\documentclass[10pt,journal,compsoc]{IEEEtran}
\usepackage{multirow}
\usepackage{booktabs} 
\usepackage[font=large,labelfont=bf]{caption}
\usepackage{subfigure}
\usepackage{tabularx}
\usepackage{multirow}
\usepackage{array}
\usepackage{amsfonts}
\usepackage{amsmath}
\usepackage{enumitem}
\usepackage{graphicx}
\ifCLASSOPTIONcompsoc
  \usepackage[nocompress]{cite}
\else
  \usepackage{cite}
\fi
\ifCLASSINFOpdf
\else
\fi
\begin{document}
%
\title{Modelling High-Order Social Relations \\for Item Recommendation}
%
%
%
%
\author{Yang Liu, Liang Chen*\thanks{*Corresponding author}, Xiangnan He, Jiaying Peng, Zibin Zheng, Jie Tang}
\IEEEtitleabstractindextext{%
\begin{abstract}
The prevalence of online social network makes it compulsory to study how social relations affect user choice. However, most existing methods leverage only first-order social relations, that is, the direct neighbors that are connected to the target user. The high-order social relations, e.g., the friends of friends, which very informative to reveal user preference, have been largely ignored. In this work, we focus on modeling the indirect influence from the high-order neighbors in social networks to improve the performance of item recommendation. 
Distinct from mainstream social recommenders that regularize the model learning with social relations, we instead propose to directly factor social relations in the predictive model, aiming at learning better user embeddings to improve recommendation. To address the challenge that high-order neighbors increase dramatically with the order size, we propose to recursively ``propagate'' embeddings along the social network, effectively injecting the influence of high-order neighbors into user representation. We conduct experiments on two real datasets of Yelp and Douban to verify our \textit{High-Order Social Recommender} (HOSR) model. Empirical results show that our HOSR significantly outperforms recent graph regularization-based recommenders NSCR and IF-BPR$^+$, and graph convolutional network-based social influence prediction model DeepInf, achieving new state-of-the-arts of the task. 
\end{abstract}

}

\maketitle

\IEEEdisplaynontitleabstractindextext

%
\IEEEpeerreviewmaketitle

\begin{figure}[t]
\centering
\includegraphics[width=0.5\textwidth]{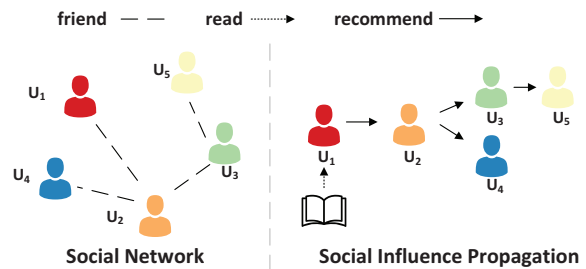} 
\caption{Toy example of the "word of mouth" influence propagation on social network. The user $U_1$ is a high-order (third-order) neighbor of $U_5$ and indirectly affects $U_5$'s choice.}\label{fig:WOM}
\end{figure}

\IEEEraisesectionheading{\section{Introduction}\label{sec:introduction}}
\IEEEPARstart{P}{ersonalized} recommendation is becoming increasingly important in online information systems in the current era of information explosion. The key success of recommendation lies in the proper use of ``collective intelligence'', i.e., a user will behave similarly with some other users. It leads to two broad categories of techniques --- collaborative filtering that discovers the similarity from user-item interactions, and content-based filtering that estimates the similarity from user/item demographic profiles~\cite{CFsurvey}. 
Nevertheless, in real-world scenarios when a user considers which items to consume, the decision choice may be affected by her friends. For example, she may ask her friends for suggestions or be attracted by products purchased by one friend. Moreover, the prevalence of online social network facilities the communication between users, which further magnifies the ``word of mouth'' influence of the social factor (see Fig.~\ref{fig:WOM} as an illustrative example). As such, to provide satisfactory recommendation service, it is important to account for the evidence in social relations when they are available to use.

Several prior efforts have been made to leverage social relations to build recommender system and verified their utility, a.k.a., the social recommendation task~\cite{ma2008sorec}. The most common solution is to design additional loss terms to encode the assumption of social influence and jointly optimize them with the recommendation objective~\cite{wang2017item,lin2018recommender,ma2011recommender,liu2018social,rafailidis2016joint,zhao2014leveraging,yu2018adaptive}. 
For example, \cite{wang2017item,lin2018recommender,ma2011recommender} assume that connected users should have similar preference and define $L_2$ smoothness terms on user embeddings to enforce the constraint; \cite{zhao2014leveraging,rafailidis2016joint} assume that the item consumed by friends should be more preferable over unobserved items and define pairwise loss terms to implement the assumption; and the recent method IF-BPR~\cite{yu2018adaptive} aggregates the two strategies of smoothness and pairwise constraint to form a unified objective function. We classify these approaches as \textit{regularization-based} methods, since the social relations are utilized to regularize the item recommendation training with additional loss terms, rather than directly enrich the predictive model. 

Despite their effectiveness, we argue that these methods are too ``implicit'' in leveraging the social relations, and the social regularization coefficients need be carefully tuned to make them work well~\cite{wang2017item}. Since the task of interest is item recommendation, a more explicit and straightforward way is to factor social relations in the predictive model that is directly optimized for recommendation. An empirical evidence is from the TrustSVD work~\cite{guo2015trustsvd}, which shows that in many cases regularization-based social recommenders can even underperform the models that are merely based on user-item interactions (e.g., SVD++~\cite{koren2008factorization}). The strong performance of TrustSVD (better than SVD++ in rating prediction) demonstrates the utility, and more importantly, the potential of factoring social relations in the predictive model. However, a limitation of TrustSVD is that it only accounts for first-order neighbors in constructing the user embedding function, making it insufficient to capture the possible ``word of mouth'' influence propagation in social network. 

Figure~\ref{fig:WOM} shows a toy example of how influence propagation from high-order neighbors can be useful for recommendation. Assume user $U_1$ likes a book and recommends it to her friend $U_2$. Although $U_2$ has not read the book due to time limitation, he still recommends it to her friends $U_3$ since he trusts $U_1$'s taste on books. This similar phenomenon happens to $U_3$, which results in an influence chain of $U_{1}\to U_{2}\to U_{3}\to U_{5}$, such that the book liked by $U_1$ indirectly affects the decision of $U_5$. 
According to some marketing research, the ``word of mouth'' effect does exist in online social networks~\cite{brown2007word} and can largely affect the sales of products~\cite{chevalier2006effect}. 
This suggests that such effect should not be ignored and motivates us to explore high-order neighbors to improve social recommender models. 

Since existing social recommenders have intensively modeled first-order social relations, it is natural to extend them by adding high-order edges in the social network (e.g., creating an edge between $U_1$ and $U_5$ in the toy example). For TrustSVD, we can extend its user embedding function by altering first-order neighbors with high-order neighbors. However, such intuitive solutions are not feasible and may not work well due to the following two challenges in modeling high-order neighbors: 
\begin{enumerate}[leftmargin=*]
    \item \textbf{Large computational cost}. Real-world social networks typically exhibit long-tail distribution on node degrees, such that a few users (hubs) have a large number of neighbors (cf. Fig.~\ref{Figure:dis}). This makes the number of high-order neighbors increase dramatically with the order size. An evidence of the increasing speed can be found in Table~\ref{table:statistic neighbors}. When we consider third-order neighbors on Douban, the average number per user is $7,413$ --- 500 times of first-order neighbor number. Thus, directly incorporating all high-order neighbors incurs large computation cost and is difficult to scale up. Although sampling strategies like random walk~\cite{yu2018adaptive} can alleviate the issue, it risks degrading the fidelity of high-order relation modeling. 
    \item \textbf{Varying importance}. While high-order social relations are useful, they may also contain noises. More importantly, they are not equally useful for different users. Intuitively, for users with many neighbors, modeling first-order relations may already be sufficient (e.g., $U_2$); while for users with few neighbors, modeling relations of more orders may be beneficial (e.g., $U_5$). To be effective in leveraging high-order relations, the method needs be able to learn varying importance for users w.r.t. different order sizes. This is difficult to achieve for existing regularization-based methods and TrustSVD. 
\end{enumerate}

\begin{table}[t]
\centering
\caption{The network density and average neighbors per user at different order sizes. }\label{table:statistic neighbors}
\begin{tabular}{|c||c|c|c|c|}
\hline
\textbf{Dataset} & \textbf{Order Size} & \textbf{Density} & \textbf{\#Neighbors/User} \\
\hline
\multirow{3}{*}{Yelp} & first & 0.15\% & 16 \\
\cline{2-4}
& second & 9.14\% & 969 \\ 
\cline{2-4}
& third & 57.16\% & 6,048 \\
\hline
\hline
\multirow{3}{*}{Douban} & first & 0.11\% & 14\\
\cline{2-4}
& second & 10.45\% & 1,332\\ 
\cline{2-4}
& third & 58.15\% &  7,413\\
\hline 
\end{tabular}\vspace{-10pt}
\end{table}

Towards the challenges in modeling high-order social relations, we propose a new predictive model for social recommendation named HOSR. Inspired by the recent developments of graph convolutional network (GCN)~\cite{kipf2016semi}, we achieve efficient modeling of high-order neighbors with the similar architecture of step-by-step message propagation. Specifically, we do one-step propagation with one GCN layer --- updating each user's embedding by aggregating the messages from her connected neighbors --- that has the complexity linear to the number of edges $|\mathcal{A}|$. By stacking $k$ such GCN layers, we do $k$-step propagations and make a user's embedding related to her $k$-order neighbors, with a linear complexity of $k|\mathcal{A}|$. 
To address the second challenge of varying importance, we design an neural attention mechanism to adaptively aggregate the user embeddings learned by different layers. Through extensive experiments on Yelp and Douban datasets, we justify the effectiveness of HOSR and provide additional insights into high-order social relation modeling for recommendation. 

To summarize, this paper makes the following contributions.
\begin{itemize}[leftmargin=*]
    \item We highlight the importance of modeling high-order social relations to capture the possible long-range influence propagation in social recommendation.
    \item We propose a new predictive model based on GCN that explicitly encodes high-order social relations into user embedding learning through multi-step message propagation. 
    
    \item We conduct experiments to show our method achieves state-of-the-art results for social recommendation, justifying the high utility of high-order social relations especially for sparse users.  
    
\end{itemize}

\section{METHODOLOGY}
Generally speaking, the proposed HOSR model consists of two components: 1) user representation learning with GCN, and 2) an attention layer which aggregates the output embedding of each GCN layer. We first formulate the problem to be solved, and then present the two components of HOSR. Finally, model optimization and time complexity are discussed.

\subsection{Task Description}
First we introduce the notation conventions. We use bold uppercase letters to denote matrices (e.g., $\textbf{U}$), bold lowercase letters to denote vectors (e.g., $\textbf{u}$), and non-bold letters to denote scalars or indices (e.g., $u$). The uppercase calligraphic symbols (e.g., $\mathcal{U}$) stand for sets. Suppose we have $n$ users and $m$ items, the interaction data between users and items are defined as an interaction matrix $\textbf{Y} = [y_{ij}]_{n\times m}$. $y_{ij} = 1$ indicates that user $i$ has a observed interaction (e.g., purchases, clicks) with item $j$. We represent the social relations between users as a user-user graph with the adjacency matrix $\textbf{A} = [a_{ii'}]_{n\times n}$. Each element $a_{ii'}=1$ indicates user $i$ and user $i'$ is connected in the social network. We now defined the  problem we study in this paper as follow.

\textbf{Input:} a user set $\mathcal{U}$, a item set $\mathcal{V}$, user-item interaction matrix $\textbf{Y}$, and user-user adjacency matrix $\textbf{A}$.

\textbf{Output:} A personalized ranking function that maps an item $\mathcal{V}$ to a real value for each user: $f_u: \mathcal{V} \rightarrow \mathbb{R}$. 

\noindent \textbf{Data sparsity problem.} Traditional matrix factorization often suffers from data sparsity problem since a large proportion of users only has a few interaction data. Thus the user preference is hard to be inferred due to the insufficient interaction data. Several efforts~\cite{wang2017item, ma2011recommender, jamali2010matrix, lin2018recommender} have been made in improving the user embedding via their first-order neighbors. However, the key challenge of integrating social relations to solve data sparsity problem is that user's social relations are very sparse as well~\cite{guo2015trustsvd}, leading to inaccurate modeling of user preference from the social perspective. In our model, user embedding is combing with the embedding propagating from user's high-order neighbors. Thus, user preference is not only inferred from her direct neighbors but also high-order neighbors, which helps to solve the social sparsity problem.

\subsection{User Representation Learning}
\subsubsection{Initial Embeddings}
We consider the recommendation problem under the representation learning framework, where each user and item is represented as an embedding vector which encodes the intrinsic features of users and items. Matrix Factorization (MF) is a well-known representation learning model that has been demonstrated effective in recommender system. The embedding of user $i$ and item $j$ is denoted by $\textbf{u}_{i}\in \mathbb{R}^{1\times d}$ and $\textbf{v}_{j}\in \mathbb{R}^{1\times d}$ respectively, where $d$ is the embedding size. Therefore, the embedding of all the users and items could be represented by two embedding matrix $\textbf{U}$ and $\textbf{V}$ where the $i$-th row of $\textbf{U}$ and $j$-th row of $\textbf{V}$ is the embedding of user $i$ and item $j$.

\subsubsection{Modeling First-Order Neighbors}
This subsection describes how to model the influence of first-order neighbors. Intuitively, a user's preference will be indirectly influenced by her social relations. As suggested in previous works~\cite{jamali2010matrix, yu2018adaptive}, a user may share similar preference with her social friends. Based on this assumption, the generation of user embedding to leverage her first-order neighbors could be implemented in two steps: 1) embedding propagation and 2) embedding aggregation. Embedding propagation explicitly models the influence between two connected users,  and embedding aggregation aggregates the embeddings propagated from user's neighbors.
\begin{figure}[t]
\vspace{-30pt}
\centering
\includegraphics[width=0.35\textwidth]{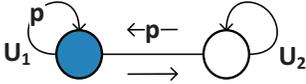} \vspace{-15pt}
\caption{The illustration of first-order embedding propagation}\label{fig:first} \vspace{-10pt}
\end{figure}

Given a user-user pair $(i, i')$, the propagation embedding $p_{ii'}$ from $i'$ to $i$ is defined as follows:
\begin{equation}\label{eq:propagation}
\textbf{p}_{ii'} = \frac{1}{\sqrt{|\mathcal{A}_{i}||\mathcal{A}_{i'}|}}\textbf{u}_{i'}\textbf{W},
\end{equation}
where $\textbf{W}\in \mathbb{R}^{d\times d^{'}}$ is the trainable weight matrix which learns useful features and transforms the embedding size from $d$ to $d^{'}$. In this paper, we set $d^{'} = d$. $\mathcal{A}_{i}$ and $\mathcal{A}_{i'}$ denote the set of first-order neighbors for user $i$ and $i'$. $1/\sqrt{|\mathcal{A}_{i}||\mathcal{A}_{i'}|}$ represents the decay factor of propagation information which defines how much user $i'$ influences the preference of user $i$. Besides considering the information from neighbors, we add self-connection $\textbf{p}_{ii}$ of user $i$ to preserve her own embedding. Figure \ref{fig:first} displays the first-order influence propagation process of two connected users. As can be seen, user $U_{1}$'s output embedding of GCN layer is generated by the origin embedding of user $U_{1}$ and her neighbor $U_{2}$. 

Following the embedding propagation, we aggregate all the embeddings from user $i$'s neighbors and user $i$'s own embedding as follows.
\begin{equation}\label{eq:agg}
\textbf{u}_{i}^{(1)} = tanh(\sum_{i'\in \mathcal{A}_{i}\cup \{i\}}\textbf{p}_{ii'}),
\end{equation}
where $\textbf{u}_{i}^{(1)}$ represents the output user $i$'s embedding of the first GCN layer. We set the nonlinear activation function as the Hyperbolic function (tanh) which empirically shows good performance. Through these two steps, we explicitly define the first-order influence propagation. 

\subsubsection{Modeling High-Order Neighbors}
\begin{figure}[t]
\centering
\includegraphics[width=0.5\textwidth]{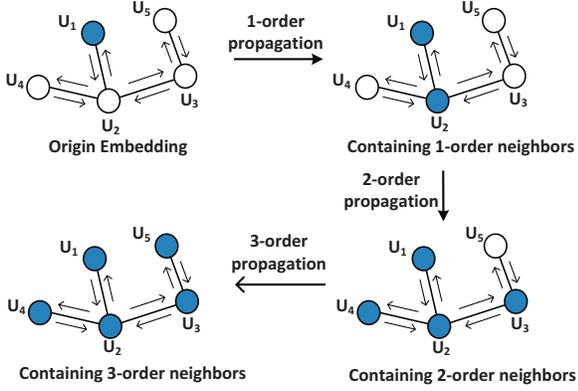}\vspace{-15pt}
\caption{The illustration of generating user embedding for user $u_{1}$ by 3rd-order influence propagation (self-connections are omitted.)}\label{fig:SIP}\vspace{-15pt}
\end{figure}

As users could be influenced by their high-order neighbors, it is crucial to model the high-order influence propagation. Through first-order influence propagation, each user's embedding contains the influence of her first-order neighbors. Therefore, through stacking more graph convolutional layers, we can assemble the features of higher-order neighbors. To be specific, by stacking $k$ GCN layers, a user $i$ can aggregate the embeddings from her $k$-order neighbors. Therefore, the embedding of user $i$ could be formulated recursively as:
\begin{equation}\label{eq:high-agg}
\textbf{u}_{i}^{(k)} = tanh(\sum_{i'\in \mathcal{A}_{i}\cup \{i\}}\textbf{p}_{ii'}^{(k)}),
\end{equation}
where the propagation embedding $\textbf{p}_{ii'}^{(k)}$ is defined as:
\begin{equation}\label{eq:high-pro}
\textbf{p}_{ii'}^{(k)} = \frac{1}{\sqrt{|\mathcal{A}_{i}||\mathcal{A}_{i'}|}}\textbf{u}_{i'}^{k-1}\textbf{W}^{(k)},
\end{equation}
where $\textbf{W}^{(k)}\in \mathbb{R}^{d\times d}$ is the trainable weighted matrix and $\textbf{u}_{i'}^{k-1}$ is the user $i'$ output embedding of the $(k-1)$-th GCN layer. As Fig. \ref{fig:SIP} shows, the embedding of user $U_{1}$ only contains her own features in the beginning. In each propagation step, user $U_{1}$ aggregates the information of higher-order neighbors. After three times influence propagation, the embedding of user $U_{1}$ consists of the information of the whole network. Therefore, we explicitly model the social influence of high-order neighbors and encode the information into the embedding of the target user.

The user embedding after $k$-layer propagation can be effectively computed by following matrix form propagation rule.
\begin{equation}\label{eq:high-agg}
\textbf{U}^{(k)} = tanh(\textbf{L}\textbf{U}^{(k-1)}\textbf{W}^{(k)})
\end{equation}
\begin{equation}
\textbf{L} = \textbf{D}^{-\frac{1}{2}}(\textbf{A} + \textbf{I})\textbf{D}^{-\frac{1}{2}},
\end{equation}
where $\textbf{W}^{(k)}\in \mathbb{R}^{d\times d}$ and $\textbf{W}^{(k-1)}\in \mathbb{R}^{d\times d}$ are the embedding matrix after $k$ and $k-1$ layer influence propagation, and $\textbf{U}^{(0)}$ is set as $U$. Each element $\textbf{L}_{ij}$ is equal to $1/\sqrt{|\mathcal{A}_{i}||\mathcal{A}_{i'}|}$ which denotes the decay factor between user $u_{i}$ and $u_{i'}$. $\textbf{A}$ is the user social relation adjacency matrix and identity matrix $\textbf{I}$ represents user's self-connections. $\textbf{D}$ represents the diagonal degree matrix where $t$-th diagonal element $\textbf{D}_{tt} = |\mathcal{A}_{t}|$.

\subsection{Attentive Layer Aggregation}
After propagating with $k$ layers, we obtain the user embedding consists of the information of $k$-th order neighbors. The user $i$'s preference towards the target item $j$ could be estimated by inner product, which is formulated as:
\begin{equation}\label{eq:pred}
\hat{y}_{ij} = \textbf{u}_{i}^{(k)}\textbf{v}_{j}^{T}
\end{equation}
However, directly utilizing the $k$-layer output $\textbf{u}_{i}^{(k)}$ cause a problem~\cite{xu2018representation}. Social networks usually consist of an expander-like core part and an almost-tree part~\cite{xu2018representation, leskovec2009community}, which represent well-connected users and the small communities respectively. As shown in Fig.\ref{Figure:dis}, most users in our experiment datasets have a few neighbors and only a few users have many neighbors. When applying too many layers, the embedding propagated to well-connected users rapidly increase. This leads to the over-smoothing problem~\cite{li2018deeper}, which means the features of each node are mixed by too many neighbors and lose locality. The over-smoothing problem make the nodes indistinguishable and hurt the model performance. On the other hand, users with a small number of neighbors need to apply more layers to aggregate sufficient information for accurate prediction. Therefore, the layer number is hard to adjust. This is why most GCN-based models~\cite{van2017graph, kipf2016semi} achieve the best performance when the layer number is one or two.

\begin{figure}[t]
\centering
\includegraphics[width=0.45\textwidth]{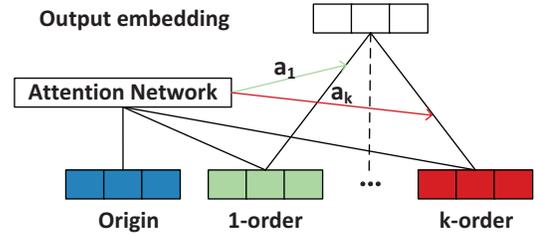} \vspace{-15pt}
\caption{The illustration of attention layer: aggregating output Embedding of each GCN layer.}\label{fig:attention}\vspace{-10pt}
\end{figure}

To solve the above problem, prior effort~\cite{xu2018representation} suggests the output embedding of different layers should be aggregated. Here we employ the attention mechanism~\cite{cao2018entive} to  evaluate the weights of neighbor orders (e.g. $k$-th order) for each node. In this paper, we use neural attention network which is widely adopted in several recommendation tasks~\cite{cao2018attentive, chen2017attentive}. Specifically, we compute a weighted sum on the output embedding of different layers, where $s_{il}$ is a learnable parameter denoting the importance of the feature learned on the $l$-th layer for user $i$. The overall neural attention network is formulated as:
\begin{equation}\label{eq:pred}
a_{il} = ReLU(\textbf{u}_{i}\textbf{P}_{u} + \textbf{u}_{i}^{(l)}\textbf{P}_{o})\textbf{h}^{T}
\end{equation}
\begin{equation}\label{eq:pred}
s_{il} = softmax(a_{il}) = \frac{exp(a_{il})}{\sum_{l^{'} = 1}^{k}exp(a_{il^{'}})},
\end{equation}
where $\textbf{P}_{u}\in \mathbb{R}^{d\times d}$ and $\textbf{P}_{o}\in \mathbb{R}^{d\times d}$ are learnable weighted matrices that project origin user embedding and output embedding of $l$-th layer to hidden layer. ReLU~\cite{nair2010rectified} is the activation function of the hidder layer. The hidden layer is then converted to attention score $a_{il}$ by vector $\textbf{h}\in \mathbb{R}^{1\times d}$. Finally, these scores are normalized by a softmax function. Then the final user embedding is calculated by:
\begin{equation}\label{eq:pred}
\begin{aligned}
\textbf{u}_{i}^{(a)} = \sum_{l}s_{ul}\textbf{u}_{i}^{(l)}
\end{aligned}
\end{equation}
Figure \ref{fig:attention} illustrates our design of attentive embedding aggregation. With such a attention network, we could evaluate the contribution of each layer to user preference.
 
\subsubsection{Model Prediction}
After aggregating the embeddings of different layers, we construct the final representations of users. As suggested by previous work~\cite{van2017graph}, combining the embeddings of user interacted items improves the model performance since the items interacted by users could be view as the user features as well. Finally, the user $i$'s preference towards the target item $j$ is given by:
\begin{equation}\label{eq:pred}
\begin{aligned}
\hat{y}_{ij} = (\textbf{u}_{i}^{(a)} + \frac{1}{\sqrt{|\mathcal{I}_{i}|}}\sum_{j'\in\mathcal{I}_{j}}\textbf{v}_{j'})\textbf{v}_{j}^{T},
\end{aligned}
\end{equation}
where $\mathcal{I}_{i}$ denotes the set of items user $i$ interacted. $1 / \sqrt{|\mathcal{I}_{i}|}$ is the decay factor. The decay factor could be replace by $1 / \sqrt{|\mathcal{I}_{i}||\mathcal{A}_{j}|}$ where $\mathcal{A}_{j}$ represents the set of users item $j$ interacted. In our experiments, we found setting decay factor to $1 / \sqrt{|\mathcal{I}_{i}|}$ performs better.

\subsection{Model Optimization}
Since we model the recommendation problem from the ranking perspective, we adopt Bayesian Personalized Ranking (BPR) loss~\cite{rendle2009bpr} for training, which is proposed to deal with the implicit feedback of users and has been widely used to optimize ranking task~\cite{zhao2014leveraging, chen2017attentive}. The assumption of BPR is that an observed interaction should be predicted with a higher score than an unobserved interaction. The loss function for training is formulated as follows.
\begin{equation}\label{eq:pred}
\begin{aligned}
loss = \sum_{(i,j^{+},j^{-})\in \mathcal{D}} -ln \sigma(\hat{y}_{ij^{+}} - \hat{y}_{ij^{-}}) + \lambda||\Theta||_{2}^{2},
\end{aligned}
\end{equation}
where $\mathcal{D} = \{(i,j^{+},j^{-})|(i, j+)\in \mathcal{D}^{+}, (u, j^{-})\in \mathcal{D}^{-}\}$ is the training set, $\mathcal{D}^{+}$ denotes the observed interaction set and $\mathcal{D}^{-}$ denotes the sampled unobserved interaction set. $\hat{y}_{ij^{+}} - \hat{y}_{ij^{-}}$ denotes the margin of predicted value between the observed interaction $(i,j+)$ and unobserved interaction $(i, j^{-})$. $\sigma(\cdot)$ is the sigmoid function. $\Theta$ is the model parameter set. 
We perform mini-batch training where we sample a batch of $(i,j,j')\in \mathcal{D}$ triples in each training epoch. Then the user embedding $\textbf{U}^{(a)}$ is generated by GCN layers and aggregated by attention layer. Lastly, we compute the loss and adopt RMSprop~\cite{hinton2012neural} algorithm to optimize our model.

Due to the strong representation power of GCN, it is prone to overfitting. To address it, we employ dropout~\cite{srivastava2014dropout}, which is an effective approach to prevent deep neural networks from overfitting. The idea is to randomly drop part of neurons during training. We use two drop techniques: \textit{graph dropout} and \textit{embedding dropout}. Following prior work~\cite{cao2018attentive, wang2017item}, embedding dropout randomly drop $p_{1}$ percent of output embedding of $k$-th layer $\textbf{u}_{i}^{(k)}$. Moreover, graph dropout randomly drop out the user's social connections with the probability $p_{2}$. Specifically, during each epoch in the training process, only $(1 - p_{2})$ of the nonzero elements appear in user adjacency matrix $\textbf{A}$. As such, in the $k$-th layer, only part of user's friends contribute the new representation. Graph dropout makes the representation more robustness against the presence or absence of social relations while embedding dropout reduces the influence of user's particular features.


\subsection{Time Complexity Analysis}
We compare the complexity of our model with TrustSVD~\cite{guo2015trustsvd}, a well-known social recommendation model considers first-order social relations and user interacted items. TrustSVD can be formulated as follows:
\begin{equation}\label{eq:trustsvd}
\begin{aligned}
\hat{y_{ij}} = (\textbf{u}_{i} + \frac{1}{\sqrt{|\mathcal{I}_{i}|}}\sum_{j'\in \mathcal{I}_{i}}q_{j'} + \frac{1}{\sqrt{|\mathcal{A}_{i}|}}\sum_{i'\in \mathcal{A}_{i}}\textbf{w}_{i'})\textbf{v}_{j}^{T},
\end{aligned}
\end{equation}
Where $\textbf{q}_{j'}\in \mathcal{R}^{1\times d}$ and $\textbf{w}_{i'}\in \mathcal{R}^{1\times d}$ denotes the implicit influence vector of item $j'$ rated by user $i$ and the user-specific embedding of user $i'$ trusted by user $i$. As we can see from equation \ref{eq:trustsvd}, the time complexity to evaluate TrustSVD is $O (|\mathcal{A}|d + |\mathcal{Y}|d)$, where $|\mathcal{A}|$ and $|\mathcal{Y}|$ denotes the number of non-zero elements in user-user social matrix \textbf{A} and user-item interaction matrix \textbf{Y}. The computation complexity of our HOSR model mainly consists of two parts: 1) influence propagation process; 2) inner product prediction. During the influence propagation process, for each layer, the matrix multiplication has computational complexity $O (|\mathcal{L}|d^{2})$, where $|\mathcal{L}|$ denotes the number of nonzero matrix in $\textbf{L}$. The computational complexity to compute the inner product is $O (|\mathcal{Y}|d)$. Therefore, the overall complexity is $O (k|\mathcal{L}|d^{2} + |\mathcal{Y}|d)$, where $k$ represents the layer number. Since $\textbf{L} = \textbf{D}^{-\frac{1}{2}}(\textbf{A}+\textbf{I})\textbf{D}^{-\frac{1}{2}}$ and multiplying $\textbf{D}^{-\frac{1}{2}}$ does not change the number of the non-zero elements, $|\mathcal{L}| = |\mathcal{I}| + |\mathcal{A}|$, where $\textbf{I}$ is identity matrix. Due to $k$ and $d$ $\ll |\mathcal{L}|$ and $|\mathcal{Y}|$, the complexity is compatible to that of TrustSVD.

\section{EXPERIMENTS}
In this section, we conduct experiments on two real-world datasets aiming to answer following research questions:
\begin{description}
	\item[RQ1] How does our proposed HOSR model perform compared with state-of-the-art social recommender approaches? Can it achieve better recommendation performance under different data sparsity level?
	\item[RQ2] How does the model performance benefit from the attention layer and high-order neighbors modeling?
	\item[RQ3] How do the two drop out strategies--embedding dropout and graph dropout affect HOSR's performance?
\end{description}

\subsection{Settings}
The statistics of datasets are displayed in Table \ref{table:statistic datasets}. As we can see, Douban dataset has more interaction than Yelp dataset. For both datasets, each user has at least one social relations. We briefly introduce the datasets we use as follows.
 \begin{itemize}[leftmargin=*]
    \item \textbf{Douban.} Douban is a famous website in China where user can rate movies, books, and songs based on her preference. Users link to others that they want to follow. As we can see from the table, this dataset has dense interaction relations. We use the book domain dataset provided by the work~\cite{hu2018leveraging}.
    \item \textbf{Yelp.} Users in yelp can rate local restaurants, gyms, bars and home services and so on. Users can post photo and reviews about these businesses as well. We use the datasets provided by the paper~\cite{shi2015semantic}. The density of social relations in Yelp is higher than Douan.
\end{itemize}

\begin{table}[t]
\centering
\caption{Dataset Statistics}\label{table:statistic datasets}\vspace{-8pt}
\begin{tabular}{|c||c|c|}
\hline
\textbf{Dataset} & \textbf{Yelp} & \textbf{Douban} \\
\hline
\hline
\textbf{\# User} & 10,580  & 12,748 \\
\textbf{\# Item} & 14,284  & 22,348 \\
\hline
\textbf{\# User-Item} & 171,102  & 785,272 \\
\textbf{\# User-User} & 169,150  & 181,890 \\
\hline
\textbf{\# User-Item density} & 0.11\%  & 0.28\%\\
\textbf{\# User-User density} & 0.15\%  & 0.11\%\\
\hline
\textbf{\# Avg. interactions} & 16.17 & 61.60\\
\textbf{\# Avg. relations} & 15.99 & 14.26\\
\hline
\end{tabular}\vspace{-10pt}
\end{table}

\begin{figure}[t] 
   \centering
   \includegraphics[width=0.23\textwidth,height=3.5cm]{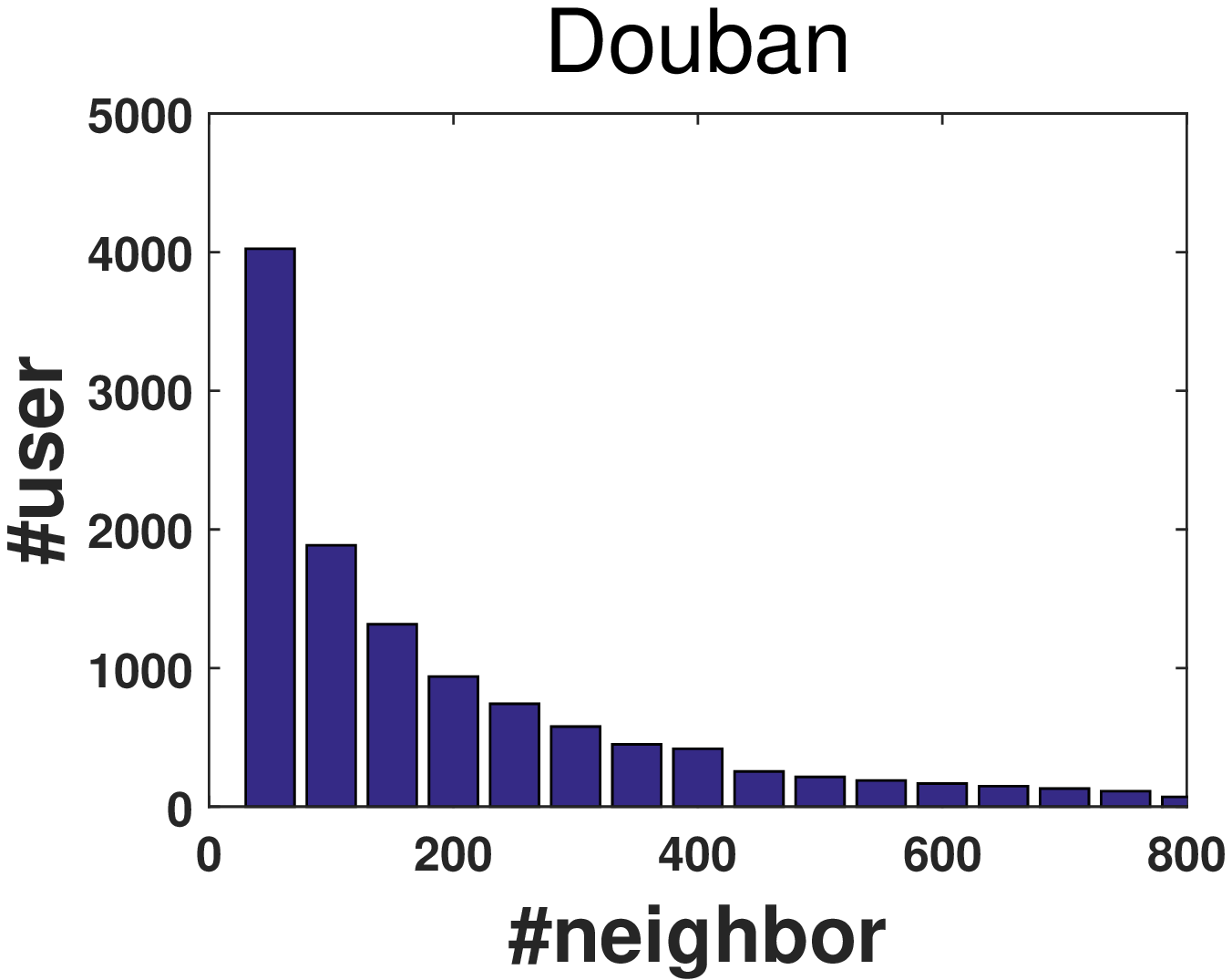} 
   \includegraphics[width=0.23\textwidth,height=3.5cm]{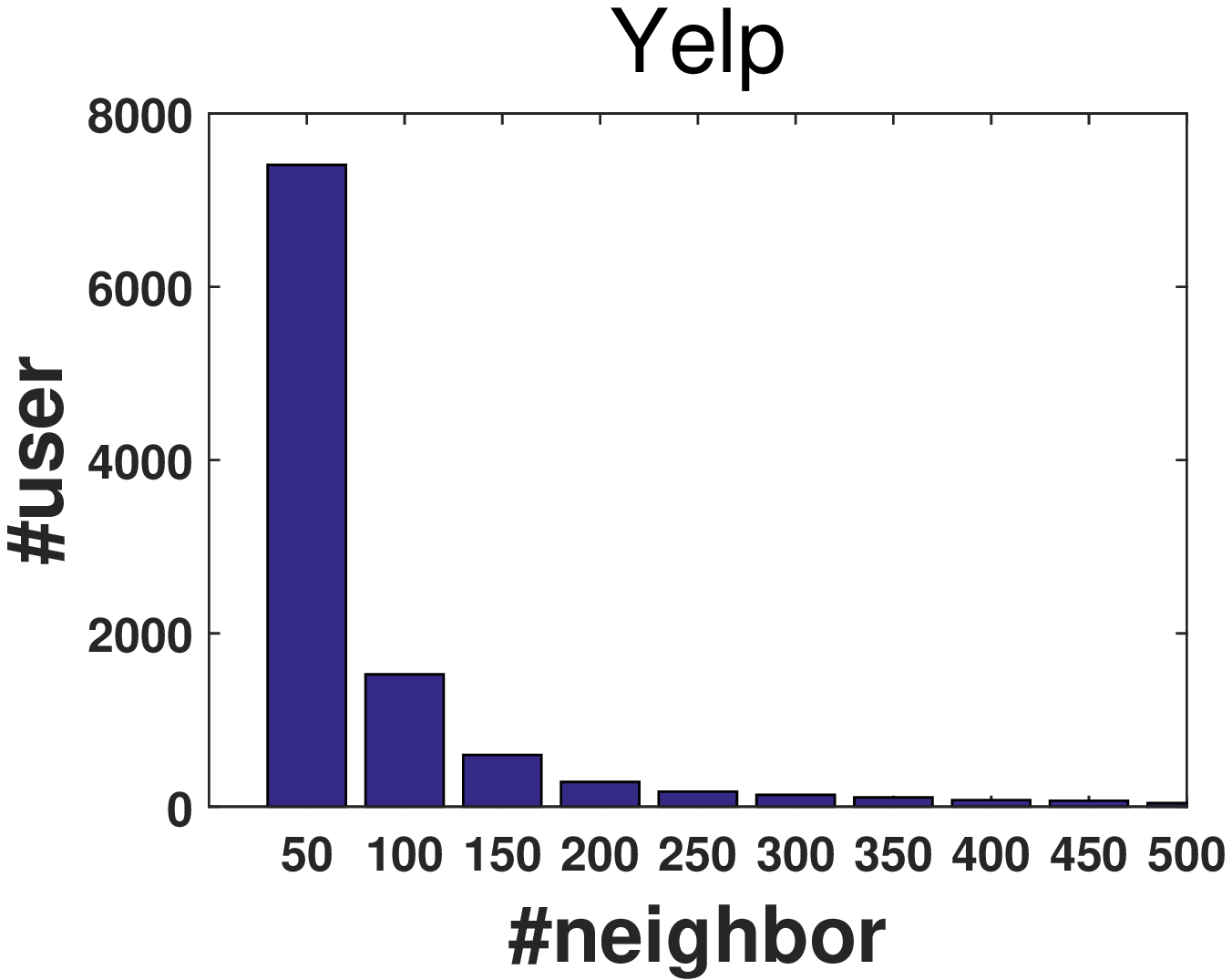}
   \caption{User distribution w.r.t. neighbor number}\label{Figure:dis}
\end{figure}

\begin{table*}[t]
\centering
\caption{Top-20 recommendation performance comparison of different methods. The last column Improv. denotes the relative improvement over the baseline: HOSR outperforms all baselines on all metrics with significance level $p$-value $<$ 0.05 (indicated by *).} \label{table:performance}\vspace{-8pt}
\begin{tabular}{|c|c|c|p{1.4cm}<{\centering}|p{1.4cm}<{\centering}|p{1.4cm}<{\centering}|p{1.4cm}<{\centering}|p{1.4cm}<{\centering}|p{1.4cm}<{\centering}|p{1.4cm}<{\centering}|c|}
\hline
\textbf{Dataset} & \textbf{Dim.} & \textbf{Metric} & \textbf{BPR} & \textbf{NCF} & \textbf{TrustSVD} & \textbf{NSCR} & \textbf{IF-BPR\textsuperscript{+}} & \textbf{DeepInf} & \textbf{HOSR} & \textbf{Improv.}\\
\hline
\hline
\multirow{8}{*}{\textbf{Douban}} & \multirow{4}{*}{\textbf{d=5}} & R@20 & 0.0659 & 0.0680 & 0.0698 & 0.0684 & 0.0663 & \textbf{0.0706} & \textbf{0.0732*} & +3.68\%\\
& & p-value & 5.24e-7 & 1.34e-2 & 1.44e-3 & 3.97e-6 & 9.33e-4 & 1.14e-10 & -- & --\\
\cline{3-11}
& & MAP@20 & 0.1981 & 0.0201 & \textbf{0.0213} & 0.0203 & 0.0194 & 0.0212 & \textbf{0.0225*} & +6.13\%\\
& & p-value & 2.04e-2 & 1.59e-4 & 4.40e-2 & 2.11e-11 & 3.49e-4 & 5.52e-3 & -- & --\\
\cline{2-11}
& \multirow{4}{*}{\textbf{d=10}} & R@20 & 0.0677 & 0.0712 & 0.0709 & 0.0708 & 0.0705 & \textbf{0.0713} & \textbf{0.0757*} & +5.63\%\\
& & p-value & 1.91e-12 & 5.13e-43 & 1.46e-5 & 4.65e-13 & 1.45e-2 & 1.45e-4 & -- & --\\
\cline{3-11}
& & MAP@20 & 0.0213 & 0.0219 & \textbf{0.0244} & 0.0230 & 0.0232 & 0.0229 & \textbf{0.0282*} & +15.57\%\\
& & p-value & 5.32e-29 & 1.65e-42 & 4.88e-19 & 1.64e-29 & 2.95e-7 & 5.66e-18 & -- & --\\
\hline
\hline
\multirow{8}{*}{\textbf{Yelp}} & \multirow{4}{*}{\textbf{d=5}} & R@20 & 0.0495 & 0.0519 & 0.0530 & 0.0522 & 0.0437 & \textbf{0.0548} & \textbf{0.0598*} & +9.12\%\\
& & p-value & 1.98e-9 & 2.93e-6 & 3.10e-5 & 2.08e-7 & 2.01e-7 & 5.49e-7 & -- & --\\
\cline{3-11}
& & MAP@20 & 0.0134 & 0.0143 & 0.0147 & 0.0148 & 0.0131 & \textbf{0.0151} & \textbf{0.0164*} & +8.61\% \\
& & p-value & 1.86e-7 & 2.23e-6 & 3.92e-4 & 6.01e-7 & 2.30e-2 & 2.60e-4 & -- & --\\
\cline{2-11}
& \multirow{4}{*}{\textbf{d=10}} & R@20 & 0.0509 & 0.0531 & \textbf{0.0570} & 0.0557 & 0.0440 & 0.0562 & \textbf{0.0697*} & +22.28\%\\
& & p-value & 7.65e-18 & 1.03e-14 & 3.50e-15 & 7.93e-12 & 1.38e-23 & 6.64e-14 & -- & --\\
\cline{3-11}
& & MAP@20 & 0.0140 & 0.0154 & 0.0151 & 0.0150 & 0.0144 & \textbf{0.0156} & \textbf{0.0202*} & +29.49\%\\
& & p-value & 2.95e-12 & 4.54e-11 & 7.42e-13 & 1.05e-11 & 1.37e-4 & 8.41e-13 & -- & --\\
\hline
\end{tabular}
\end{table*}

To gain insights into the data with respect to user's social relations, the user distributions with respect to the number of social neighbors are shown in Fig.\ref{Figure:dis}. As we can see, both datasets show a long-tail distribution -- most users have few neighbors and only a small proportion of users have many neighbors. This observation highlight the social sparsity challenge faced by social recommendation.

For both datasets, $80\%$ of the whole data are randomly selected for training and the remaining $20\%$ are for testing. To evaluate the top-K recommendation performance of the models, we adopt a relevance-based metric -- \textit{Recall@K} and a ranking-based metric -- \textit{MAP@K}(Mean Average Precision). In our experiments, all the items that a user hasn't consume are treated as negative items. Recall@K measures the number of positive items that present within the top-K recommendation list divided by all the positive items. MAP@K considers the ranking positions of the positive items within the top-K of the recommendation list. Large values of R@20 and MAP@20 indicate better performance.

\begin{table*}[t]
\centering
\caption{The effect of attention layer and high-order neighbors modeling: the base model represents HOSR without attention layer and the average model assign the output of each layer the same weight.}\vspace{-8pt} \label{table:attention}
\begin{tabular}{|c||c|c|c|c|c|c|c|c|c|c|c|c|}
\hline
\multirow{3}{*}{\textbf{Model}} & \multicolumn{6}{c|}{\textbf{Douban}}  & \multicolumn{6}{c|}{\textbf{Yelp}} \\
\cline{2-13}
& \multicolumn{2}{c|}{\textbf{Base}} & \multicolumn{2}{c|}{\textbf{Average}} & \multicolumn{2}{c|}{\textbf{Attention}} & \multicolumn{2}{c|}{\textbf{Base}} & \multicolumn{2}{c|}{\textbf{Average}} & \multicolumn{2}{c|}{\textbf{Attention}} \\
\cline{2-13}
& R@20 & MAP@20 & R@20 & MAP@20 & R@20 & MAP@20 & R@20 & MAP@20 & R@20 & MAP@20 & R@20 & MAP@20\\
\hline
\textbf{HOSR-1} & 0.0716 & 0.0229 & -- & -- & -- & -- & 0.0636 & 0.0176 & -- & -- & -- & --\\
\hline
\textbf{HOSR-2} & 0.0747 & 0.0241 & 0.0734 & 0.0231 & 0.0748 & 0.0249 & 0.0657 & 0.0183 & 0.0637 & 0.0177 & 0.0653 & 0.0181\\
\hline
\textbf{HOSR-3} & 0.0726 & 0.0239 & 0.0743 & 0.0247 & 0.0757 & \textbf{0.0282} & 0.0640 & 0.0179 & 0.0674 & 0.0187 & \textbf{0.0697} & \textbf{0.0202}\\
\hline
\textbf{HOSR-4} & 0.0720 & 0.0236 & 0.0759 & 0.0255 & \textbf{0.0773} & 0.0260 & 0.0630 & 0.0169 & 0.0658 & 0.0173 & 0.0641 & 0.0179\\
\hline
\end{tabular}
\end{table*}

We compare our proposed model with the following baselines.
\begin{itemize}[leftmargin=*]
    \item \textbf{BPR}~\cite{rendle2009bpr}: This is the matrix factorization model that minimizes the Bayesian personalized ranking (BPR) loss. This model only utilizes the interaction data between users and items.
    \item \textbf{NCF}~\cite{he2017neural}: This method is a state-of-the-art neural CF model which combines element-wise and hidden layers of the concatenation of user and item embedding to capture their high-order interactions.
    \item \textbf{NSCR}~\cite{wang2017item}: This model adopts a deep neural network in modeling the user-item interactions and uses two social constraints (smoothness and fitting) to learn the user representation.
    \item \textbf{IF-BPR}~\cite{yu2018adaptive}: IF-BPR explicitly defines several paths to identify the user's possible friends based on heterogeneous information network. It divides all the items into five classes and learns a ranking function through a predefined order of these five class.
    \item \textbf{TrustSVD}~\cite{guo2015trustsvd}: TrustSVD jointly models the first order user-user social relations and user-item interaction based on SVD++ framework. In our experiments, we optimize TrustSVD by BPR loss.
    \item \textbf{DeepInf}~\cite{qiu2018deepinf}: DeepInf is a social influence prediction model. It conducts the random walk with restart strategy to sample fix-size neighbors for each user. They utilizes GCN to generate user embedding and predict the social influence. We adapt DeepInf to the social recommendation task. We first learn user embedding by DeepInf and then employ the dot product between user and item embedding to predict the user preference.
\end{itemize}

\noindent\textbf{Hyper-parameter Settings.} We implement BPR, NCF, NSCR, DeepInf, TrustSVD and our model based on Pytorch, which is optimized with the RMSprop optimizer. We use the code of IF-BPR provided by the authors\footnote{https://github.com/Coder-Yu/RecQ}. The batch size is fixed to 512 for all models. The learning rate is tuned in [0.0001, 0.0005, 0.001, 0.005]. The coefficient of $L_{2}$ regularization is ranged in [0.0001, 0.001, 0.01, 0.1]. We apply embedding dropout for NCF, NSCR and DeepInf, where the dropout ratio is ranged in [0, 0.1, 0.2, ... , 0.8]. We use 3 layer structure for NCF, NSCR, DeepInf and our model, where the dimension of each layer keeps the same. For DeepInf, the sample size is set to 50 and return probability is set to 0.5. We set the embedding dropout of 0 and graph dropout of 0.2 for our HOSR model.

\subsection{Performance Comparison (RQ1)}
\subsubsection{Overall Comparison.} Top-K recommendation performance of HOSR and other state-of-the-art models are reported in Table \ref{table:performance}. The model performances are compared under the embedding size of 5 and 20. The parameter K is set to 20. Based on the experimental results, we have the following observations.
\begin{itemize}[leftmargin=*]
    \item BPR performs worst in all cases. This indicates that directly optimize user embedding through inner product is insufficient to capture the user preference towards items. NCF performs better than BPR, indicating the effectivenee of applying deep neural network to learn the nonlinear user-item feature interactions. As we could see from the table, social recommendation models outperform non-social models, which indicates incorporating social relations is beneficial to model performance.
    \item Compared to BPR and NCF, TrustSVD and NSCR achieve better performance. This verifies integrating the information of first-order neighbors could improve the model performance.  Furthermore, TrustSVD achieves the best performance in some cases, demonstrating the importance of explicit modeling of both social and interaction relations.
    \item Although IF-BPR models the influence of high-order neighbors, it does not outperform NSCR and TrustSVD. The reason might be IF-BPR learns the high-order relations in implicit ways and does not assign a different weight for the selected neighbors, which might introduce noise.
    Moreover, although the same paths are used in both datasets, their performance is different. IF-BPR performs better in Douban than Yelp, demonstrating the same defined paths may not generalize across datasets. One possible solution is to select the useful paths according to datasets. However, this process is time-consuming and hard to tune.
    \item DeepInf achieves better performance than other baselines in most cases. Such improvement verifies the effectiveness of high-order neighbors modeling. Though both DeepInf and HOSR consider high-order neighbors, HOSR performs better than DeepInf, indicating our method better models the social recommendation problem.
    \item HOSR achieves the best performance under all circumstances. In particular, HOSR achieves 5.63\%, 15.57\% and 24.02\%, 22.28\% performance improvement against the strongest baseline w.r.t R@20 and MAP@20 when the embedding size is 10 in Douban and Yelp, demonstrating the effectiveness of our model. Compared to TrustSVD, despite both HOSR and TrustSVD consider the user-item interaction and user social relations, HOSR outperforms TrustSVD, verifying that high-order neighbors modeling is beneficial to model performance.
\end{itemize}

\begin{figure*}[t]
   \centering
   \includegraphics[width=0.24\textwidth,height=3.5cm]{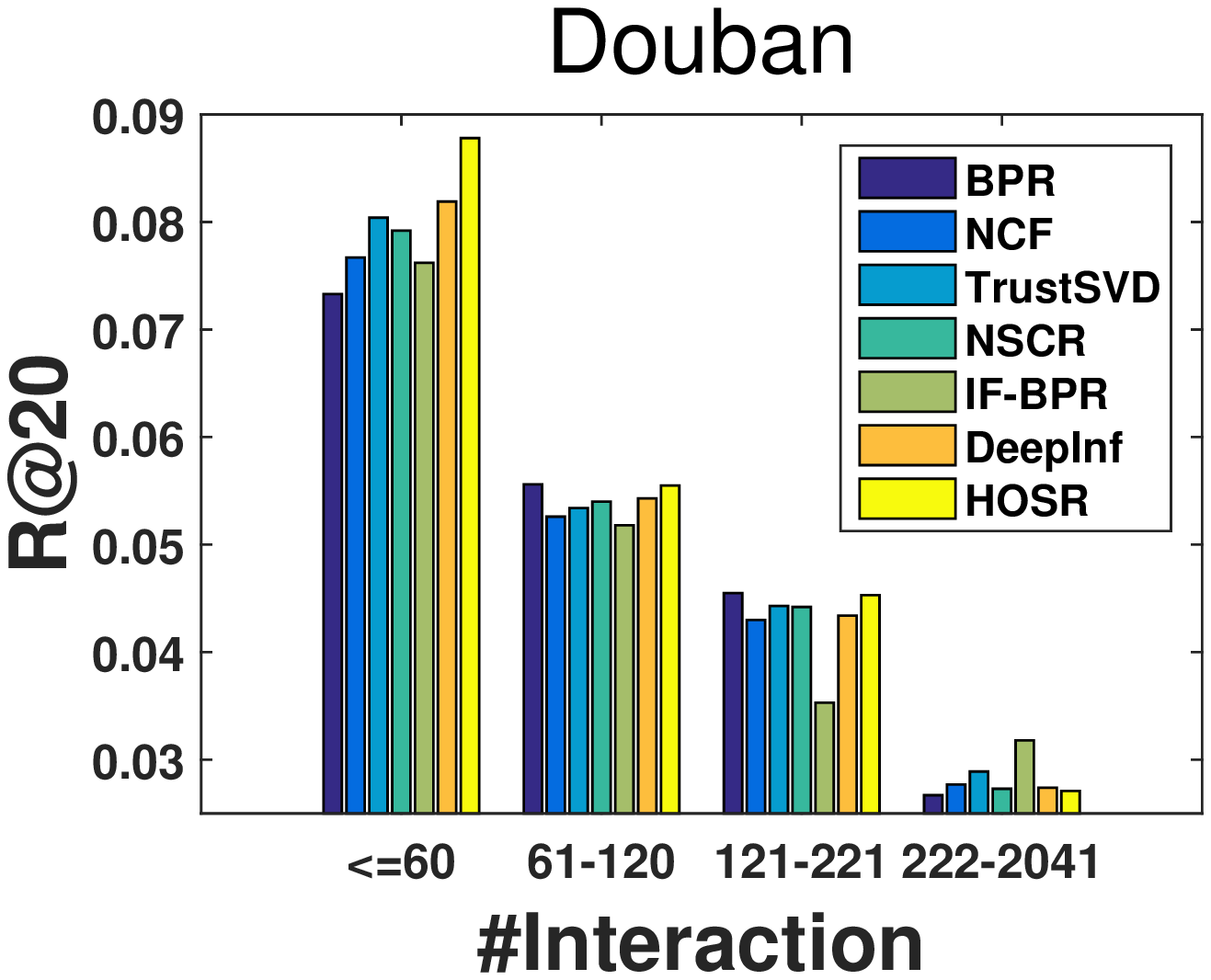}
   \includegraphics[width=0.24\textwidth,height=3.5cm]{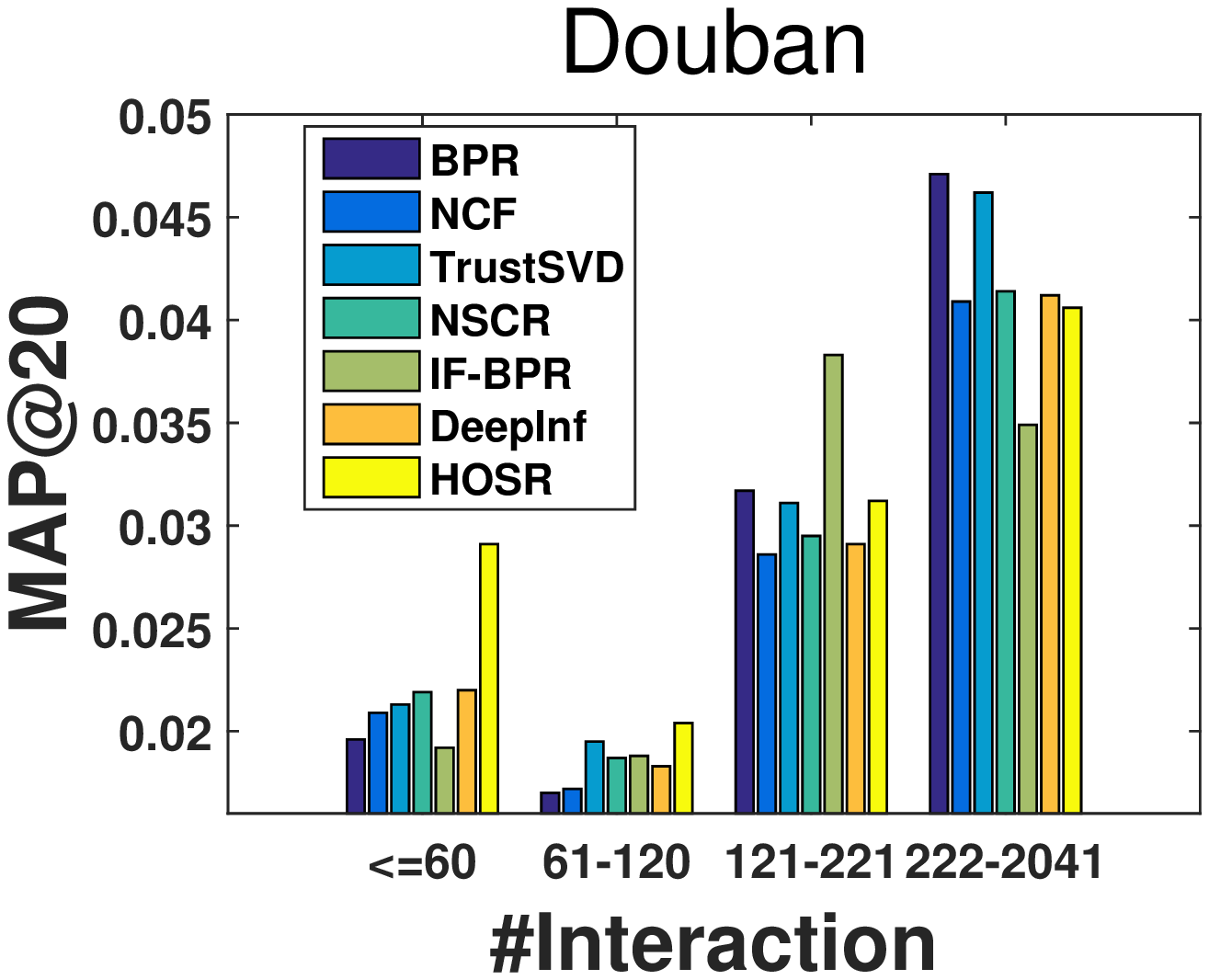}
   \includegraphics[width=0.24\textwidth,height=3.5cm]{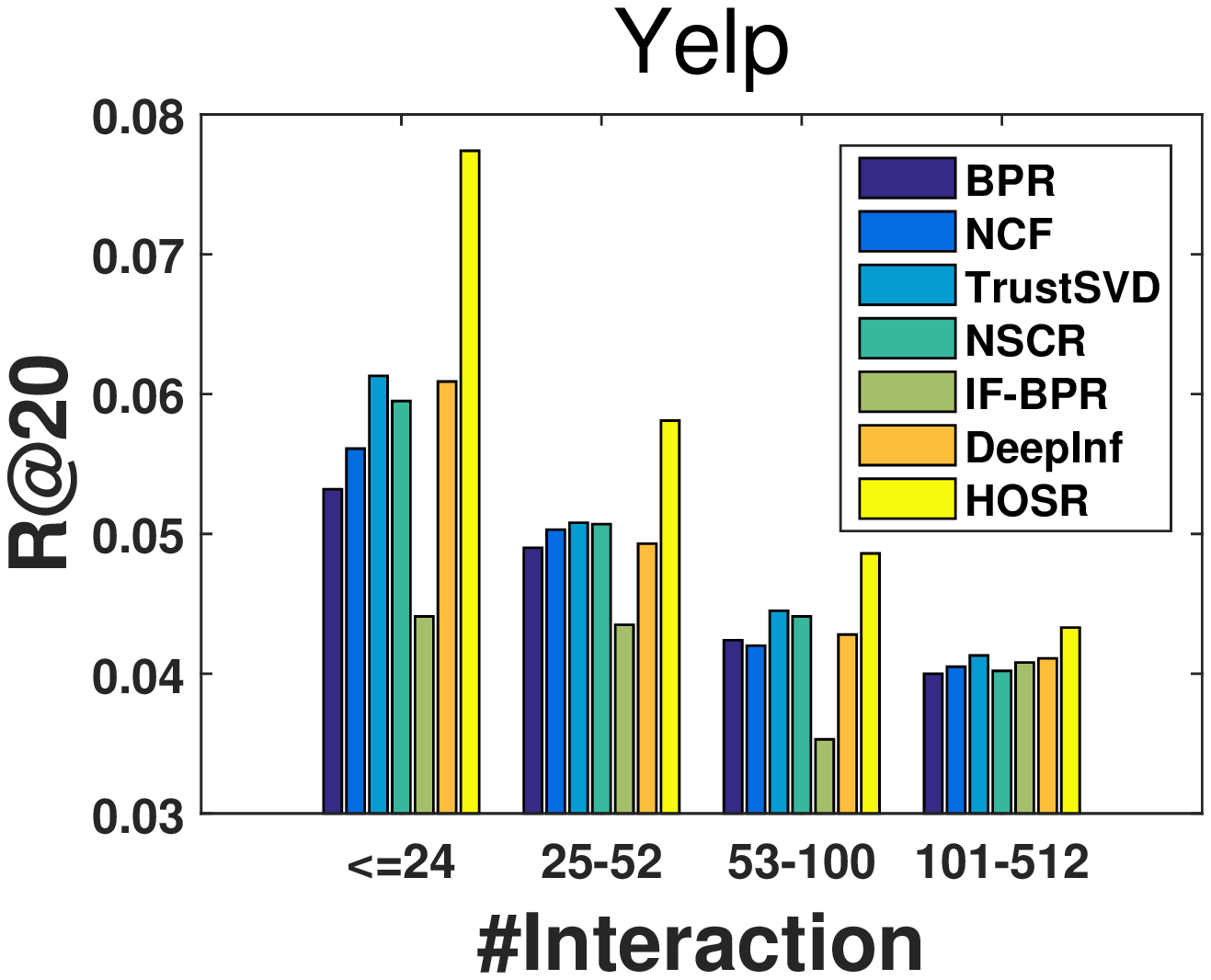}
   \includegraphics[width=0.24\textwidth,height=3.5cm]{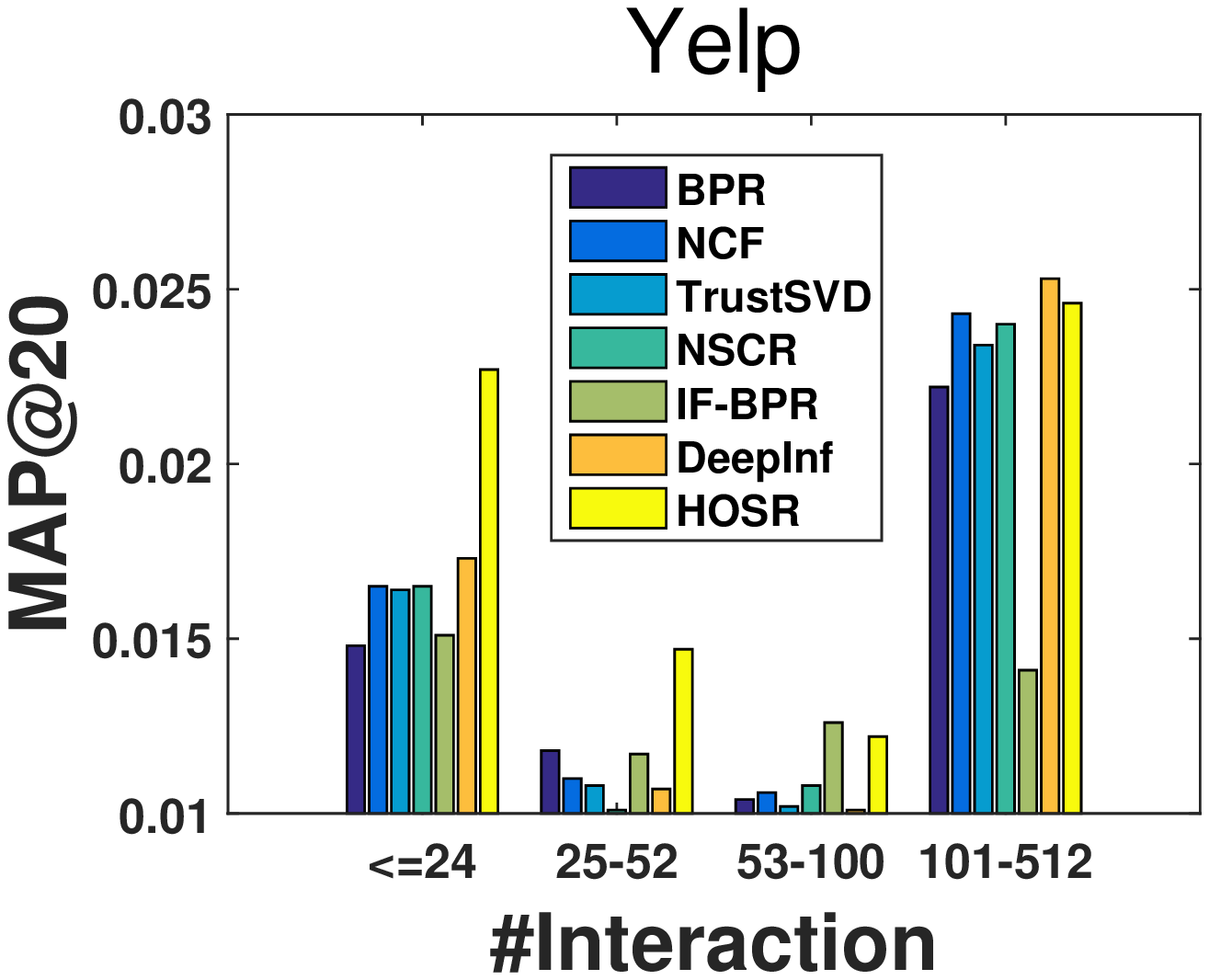}
   \caption{Experimental results of the different sparsity level user groups. The improvement is more significant when the interactions of users are sparser.} \label{fig:group}
\end{figure*}

\subsubsection{Performance Comparison w.r.t Interaction Sparse Levels.} To investigate the effect of data sparsity, we divide the test users into four groups based on their interaction number in the training set. Each group has the same total interactions. As could be seen in Fig.\ref{fig:group}, the interaction number per user of these four groups is $\leq$60, 61-120, 121-221, 222-2041 in Douban and $\leq$24, 25-52, 53-100, 101-512 in Yelp respectively. We report the experimental results on different user groups and all test users w.r.t. Recall@20 and MAP@20 in Fig.\ref{fig:group}. According to the table, we have the following observations:
\begin{itemize}[leftmargin=*]
    \item By analyzing all the figures in Fig.\ref{fig:group}, compared to other baselines, we observe that HOSR achieves better performance over the $<=60$ and $61-120$ group in Douban and $<=24$, $25-52$ and $53-100$ group in Yelp respectively. Moreover, the improvement is more significant when the user group is sparser, demonstrating that modeling high-order social influence is beneficial to the relatively sparse users.
    \item For users that consumed a large number of items (the fourth group), the performance does not improve compared to other methods. The reason might be modeling the user-item interactions is sufficient to capture the user preference for these active users. As the figure shows, the performance of BPR of fourth group users is comparable to other methods. This also demonstrates the importance of modeling the preference of users with few interaction data.
\end{itemize}

\begin{figure*}[htbp] 
  \centering
  \includegraphics[width=0.24\textwidth,height=3.5cm]{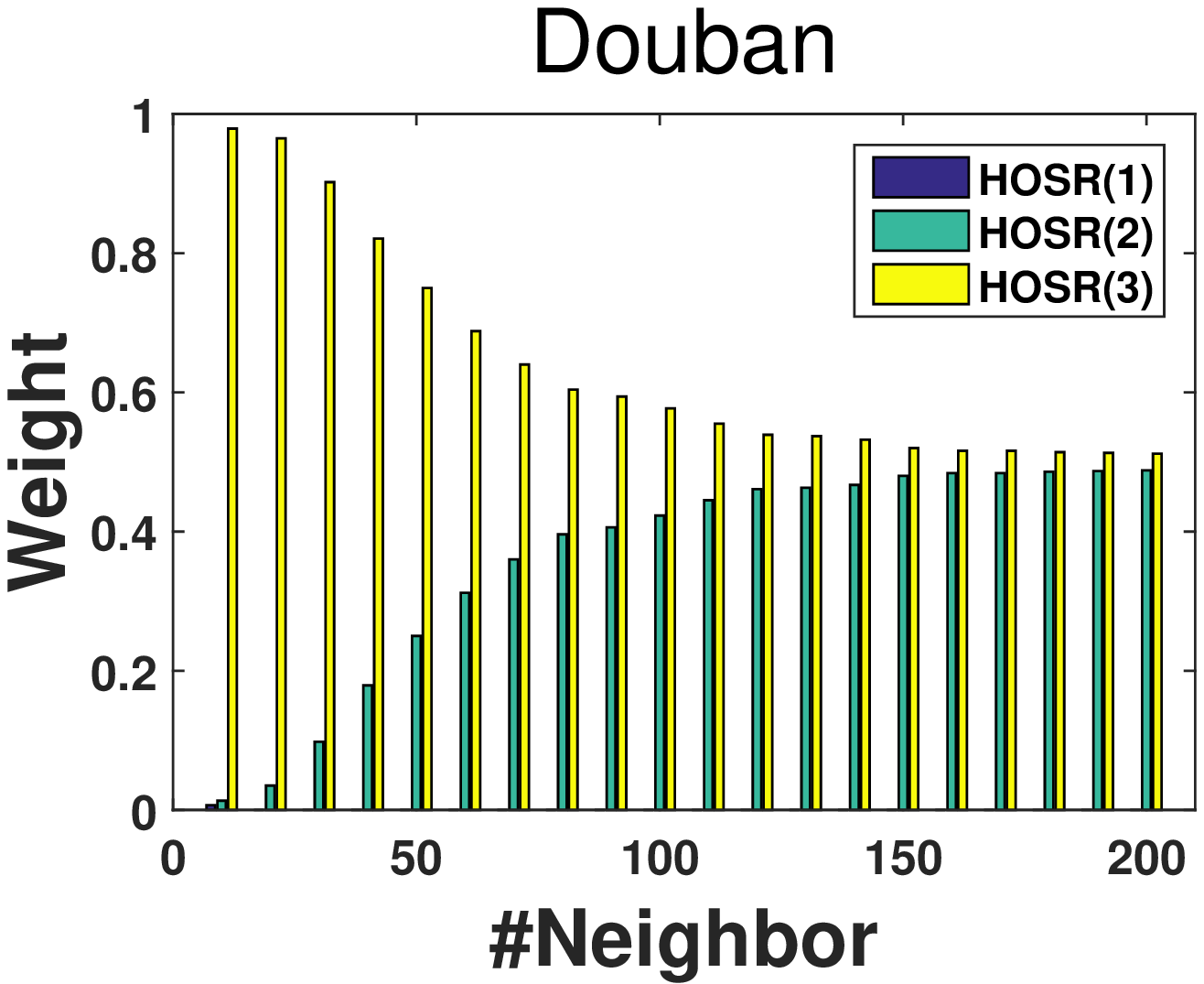} 
  \includegraphics[width=0.24\textwidth,height=3.5cm]{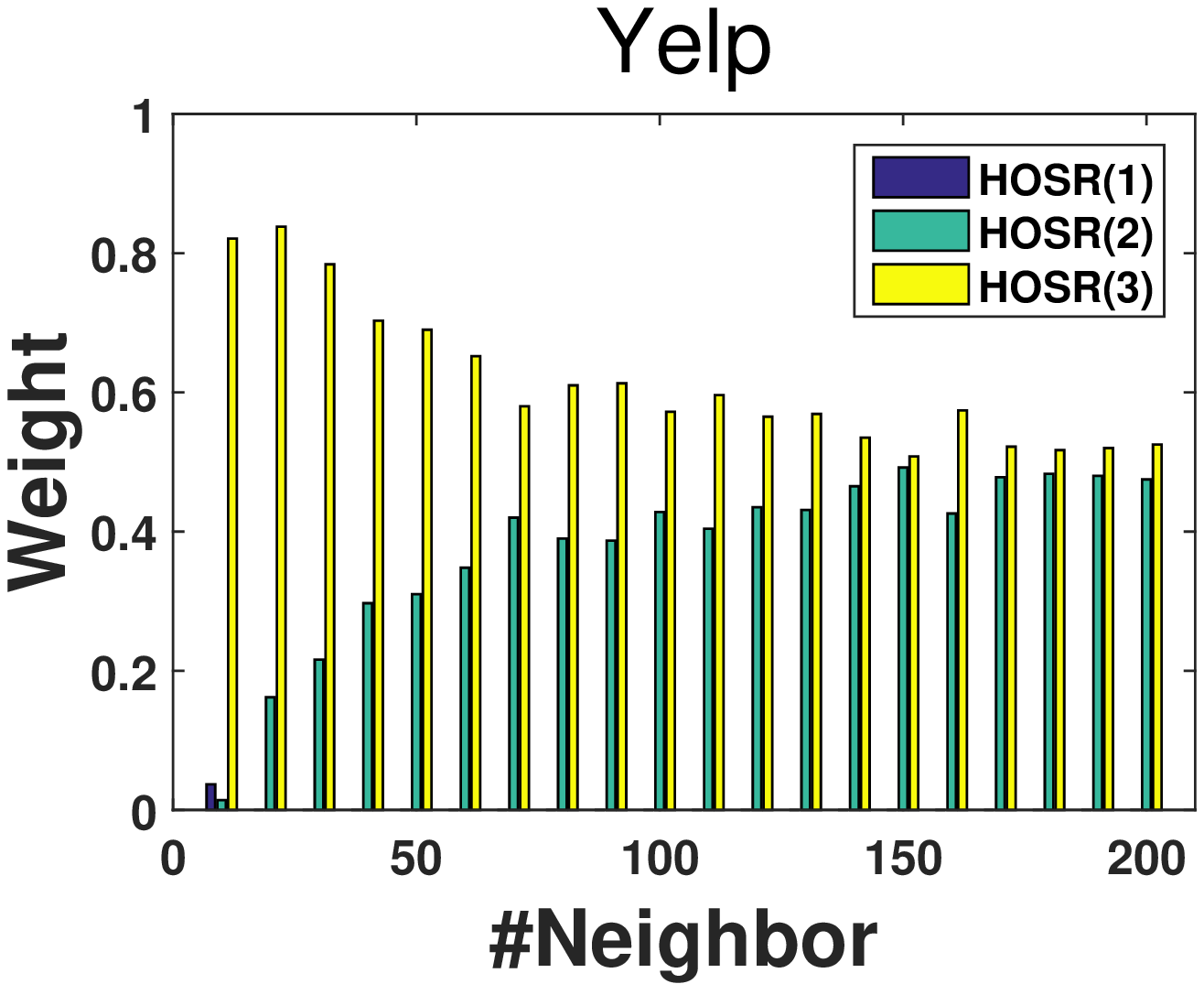}
  \includegraphics[width=0.24\textwidth,height=3.5cm]{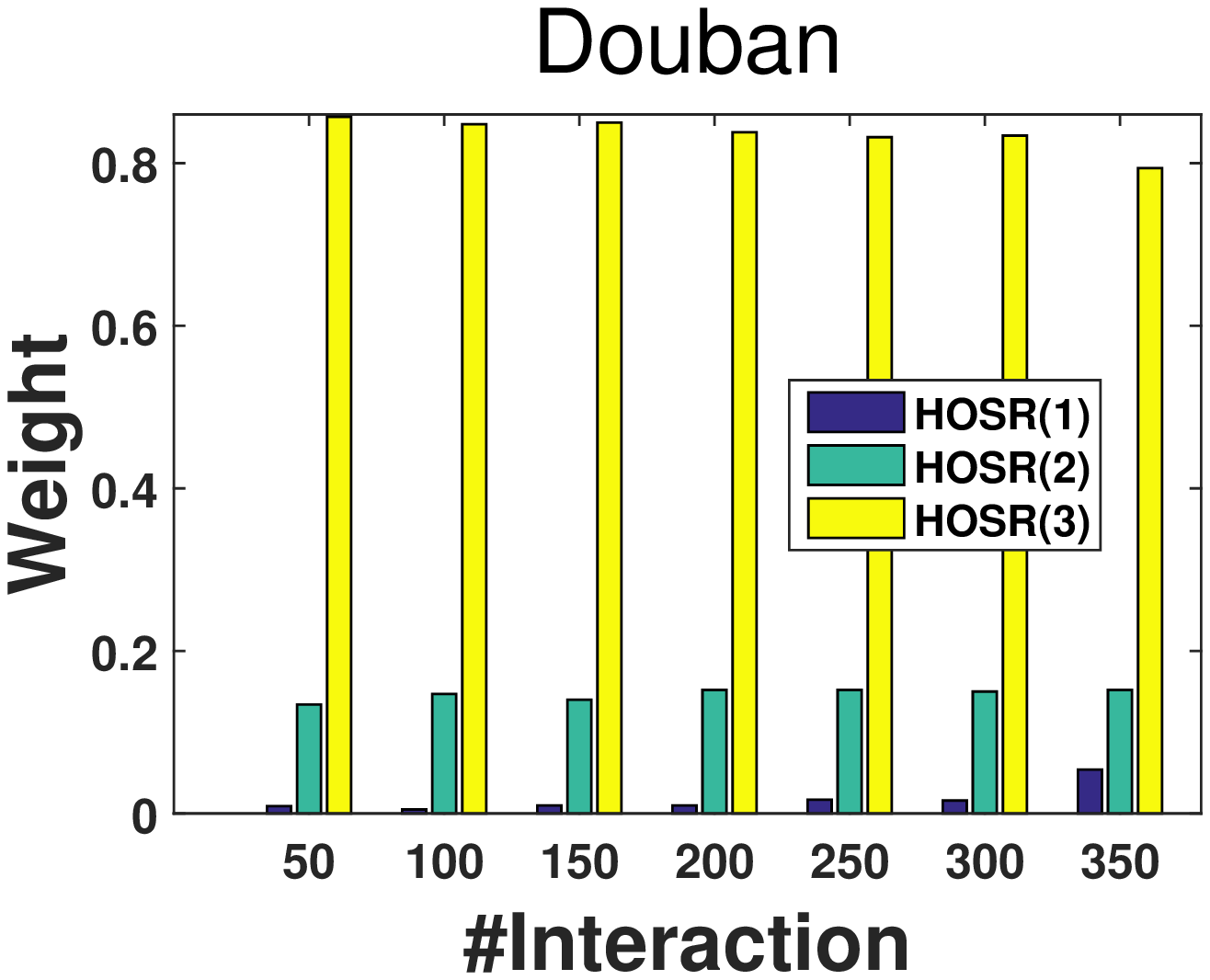} 
  \includegraphics[width=0.24\textwidth,height=3.5cm]{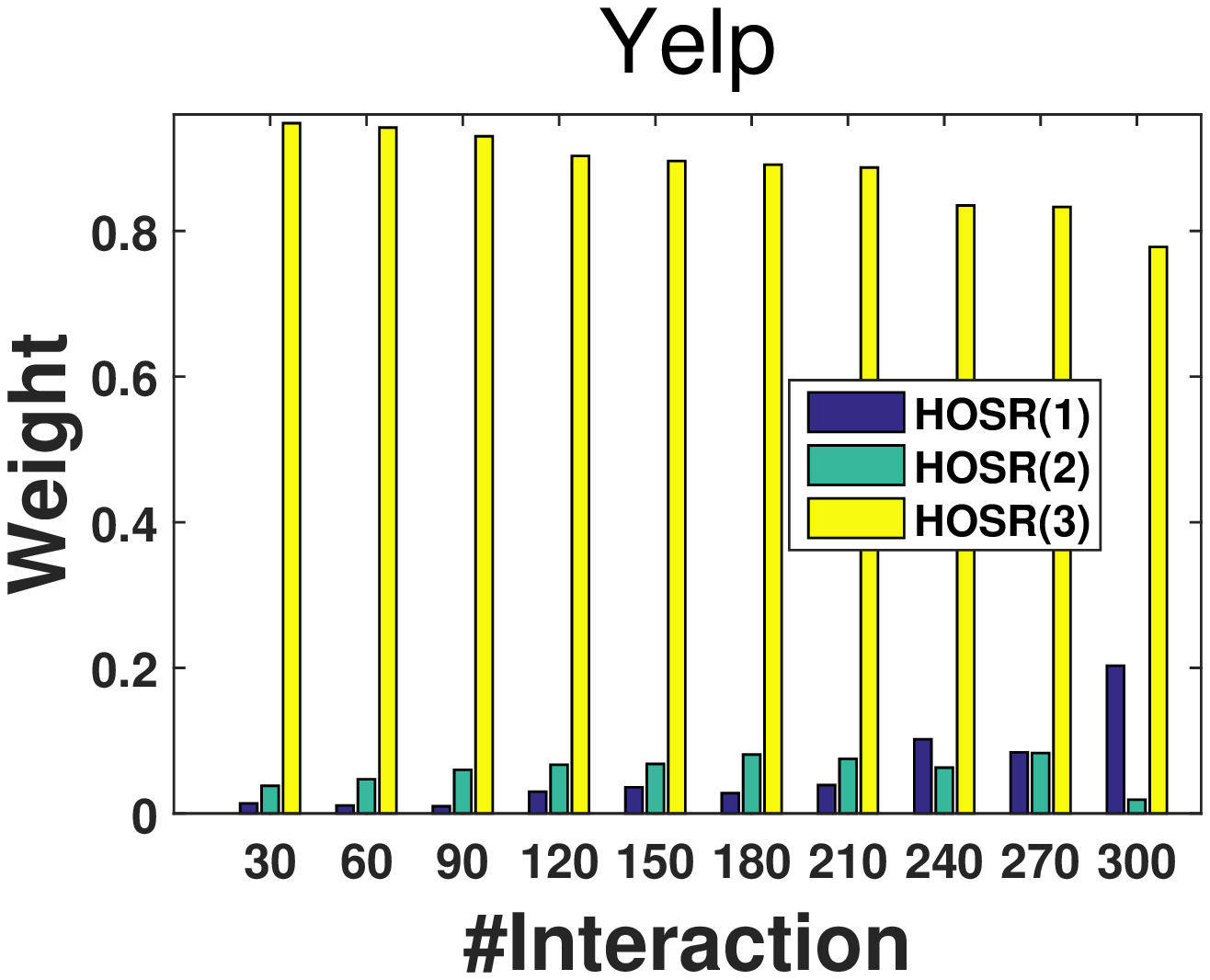} 
  \caption{Attention weight of different layers w.r.t \#neighbor and \#Interaction. The attention weights decrease with the increase of both the number of social relations and item interactions of users.}\label{fig:attention_dis}
\end{figure*}

\subsection{Component Analysis (RQ2)}\label{sec:attention}

To investigate the effect of attention layer and high-order neighbors modeling, we compare the HOSR with two model variants -- the \textit{base model} which removes the attention layer and the \textit{average model} which uses the average of the output of each layer as user embedding. The layer number ranges from 1 to 4. We use HOSR-$k$ to denote the model with $k$ layers. The attention mechanism is used only when layer number is greater than 1. The experimental results are shown in Table \ref{table:attention}. Based on the experimental results, we have the following observations:
\begin{itemize}[leftmargin=*]
    \item As shown in Table \ref{table:attention}, compared to the base and average model, employing attention layer to aggregate the output of different layers achieves the best performance. The reason is attention mechanism flexibly learns different weights for different users. Thus the varying importance of different layers for different users is successfully captured.
    \item The best performance of \textit{base model} is achieved when the layer number is two. The reason might be stacking more layers aggregates too much information, leading to over-smoothing problem, further limits the model performance. 
    \item When stacking more layers, in most cases, the performance of the attention model first increases, then the performance decreases. The reason is when the layer number is small, the information aggregated is not sufficient to model the user preference. Therefore stacking more layers enhances the model performance. After a threshold, adding more layers may lead to overfitting problem and introduce noise as well.
\end{itemize}

To further understand how attention layer aggregates the output of different layer and how attention weights distribute, we visualize the attention weight with respect to the number of social neighbors and the number of interactions. The results is displayed in Fig.\ref{fig:attention_dis}. We have the following observations:
\begin{itemize}[leftmargin=*]
    \item As the figures display, in all the cases, the weight of the first layer is small. This might because stacking more layers not only leverages the information of neighbors but also selects useful features. Therefore, the output of the second and third layer is more important.
    \item As the figures with respect to the number of neighbors (i.e., the left two figures in Fig.\ref{fig:attention_dis}) shows, when the number of neighbors is small, the attention weight of the last layer is extremely high, demonstrating the importance of modeling high-order neighbors for users with sparse social relations. When the number of neighbors increases, the attention weight of the last layer decreases while the attention weight of the second layer increases. The reason is modeling high-order neighbors introduces noise for users when user's neighbors increase therefore the importance of last layer decreases. This also implies our model could balance the importance of different layer output.
    \item As could be seen in figures with respect to the number of interactions (i.e., the right two figures in Fig.\ref{fig:attention_dis}), for both datasets, the attention weights of the last layer are much higher than other layers, demonstrating the importance of high-order neighbors in capturing the user preference. when the number of interaction decreases, the attention weight of the last layer increases, verifying for sparse users, the output of the last layer is more important than active users.
\end{itemize}

\begin{figure*}[htbp]
   \centering
   \includegraphics[width=0.24\textwidth,height=3.5cm]{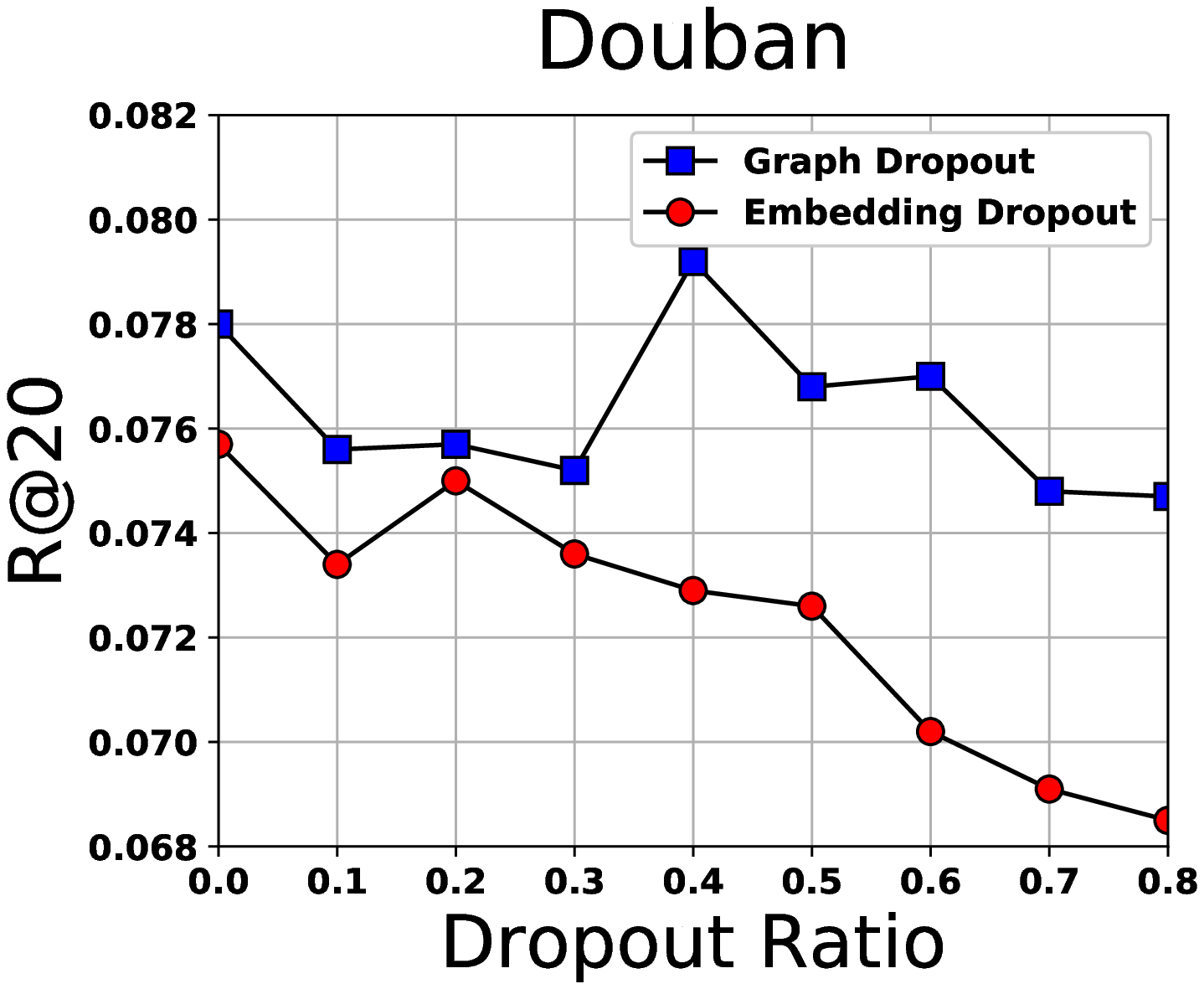}
   \includegraphics[width=0.24\textwidth,height=3.5cm]{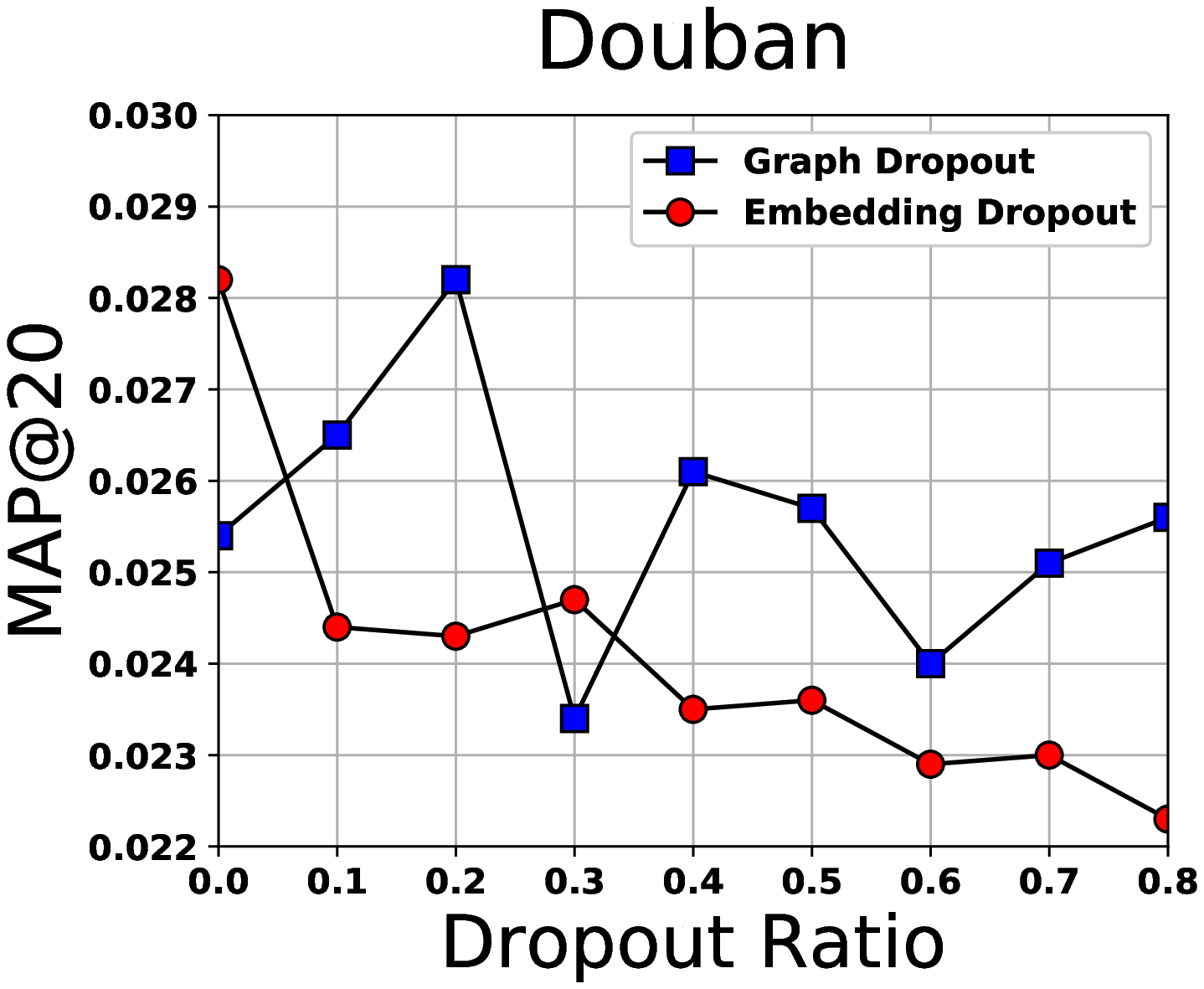}
   \includegraphics[width=0.24\textwidth,height=3.5cm]{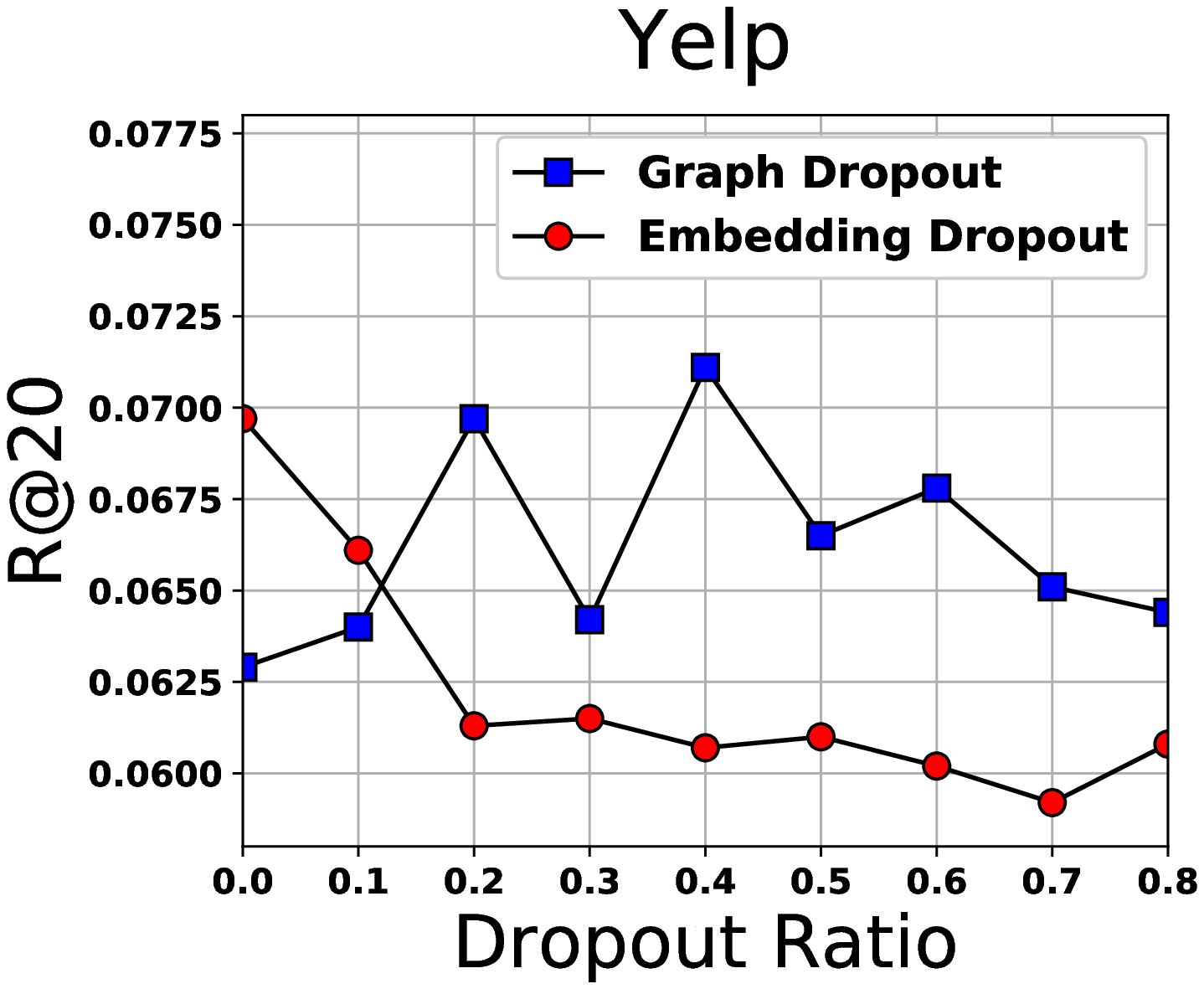}
   \includegraphics[width=0.24\textwidth,height=3.5cm]{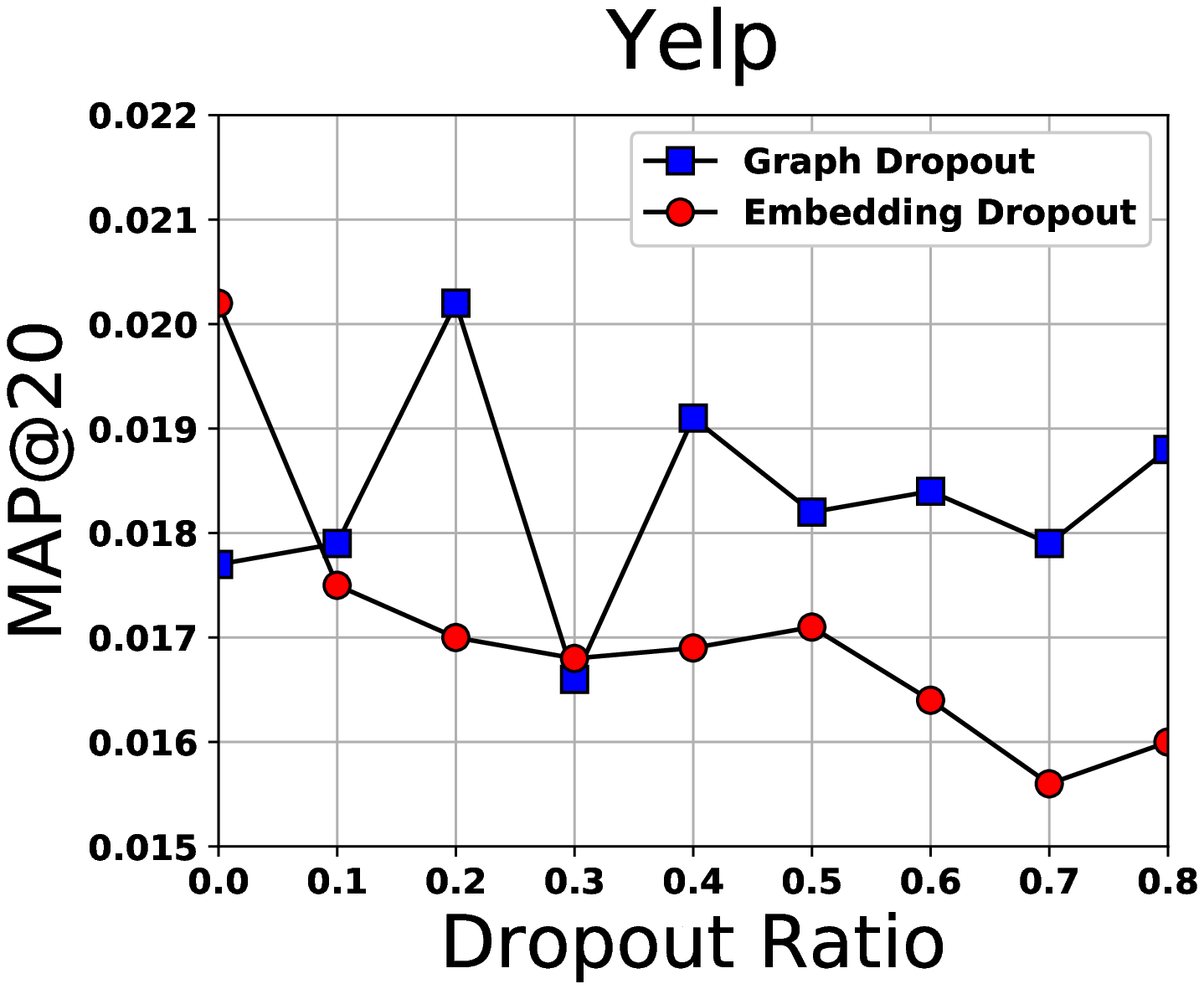}
   \caption{Effect of graph dropout and embedding dropout}\label{fig:dropout}
\end{figure*}

\subsection{Effect of dropout (RQ3)}
In our work, graph dropout and embedding dropout is employed to prevent overfitting in our model. Figure \ref{fig:dropout} displays the effect of embedding dropout ratio $p_{1}$ and graph dropout ratio $p_{2}$ on both datasets.

As could be seen from the Fig.\ref{fig:dropout}, applying embedding dropout does not enhance the model performance. The reason is the output of each layer not only consist of the information of the user itself but also the information of her neighbors. Thus embedding dropout might easily lose information. Compare to embedding dropout, graph dropout offers better performance. Especially, setting $p_{2}$ as 0.4 leads to the highest R@20 and $p_{2}$ as 0.2 leads to the highest MAP@20 on Yelp. One reason might be dropping out part of social relations from users makes the representations more robust against noises when stacking more layers.

\section{RELATED WORK}
In this section, we highlight some related works to this paper, including the works of social recommendation, graph representation learning and attention mechanism respectively.
\subsection{Social Recommendation}
With the rapid development of online social platforms, social connection has been widely studied and leveraged to enhance the recommendation performance~\cite{liu2018social, DBLP:conf/sigir/SunW018}. We categorize previous social recommendation models related to our work into three types. The first type of methods consider social network as a regularization~\cite{jamali2010matrix, wang2017item, tang2016recommendation, lin2018recommender, ma2011recommender}. SocialMF~\cite{jamali2010matrix} proposes to constraint user's latent vector to be close to the weighted average of his social neighbors. SoDimRec~\cite{tang2016recommendation} considers the heterogeneity of social relations and weak dependency connections as a regularization. CSR~\cite{lin2018recommender} propose Characterized Social Regularization to model user's various similarities with different friends.

The second type of methods first divide all the items into different sets and then manually define the ranking relations between these sets\cite{zhao2014leveraging, yu2018adaptive}. SBPR~\cite{zhao2014leveraging} divides the item set into positive items, social items and negative items and claims all the items should follow the ranking: positive items $>$ social items $>$ negative items. IF-BPR~\cite{yu2018adaptive} further divides the item set into five groups: positive items, joint social items, positive social items, negative social items and non-consumed items based on heterogeneous information network.

The third type of methods combine user's embedding with their social neighbors'~\cite{chaney2015probabilistic, ma2009learning, guo2015trustsvd}. RSTE~\cite{ma2009learning} linearly combines the user representation with those of the user's social neighbors. SPF~\cite{chaney2015probabilistic} model the prediction as  the combination of the product of latent vectors and the ratings from social neighbors under the probabilistic Poisson factorization framework.

Despite the success of these models, most of them only consider implicitly model the connections of user's first order neighbors, which leads to a suboptimal prediction. Note that although DeepInf also models the high-order neighbors, they aim to predict the social influence. Moreover, although DeepInf~\cite{qiu2018deepinf} samples fix-size neighbors to balance the neighbor number among users, it might lose information as well. Besides, it is hard to tune the sample size and the return probability of the random walk.  

\subsection{Graph Neural Network}
Graph neural network (GNN) ~\cite{scarselli2009graph} extends the neural network to processing the graph data. GNN aims to learn an embedding for each node which contains the information of neighbors. Graph convolutional network (GCN)~\cite{kipf2016semi} is a special type of GNN which applies convolutional operations on graph data. The authors propose the spectral graph convolution which could directly operate on graph data. Moreover, GraphSAGE~\cite{HamiltonYL17} learns a function to aggregating node's local neighbors' feature by sampling. Graph attention network(GAT)~\cite{velivckovic2017graph} employs attention mechanism to aggregate the neighbors' feature vectors. GNN-based models have achieved the state-of-the-art performance on several tasks, such as social influence prediction~\cite{qiu2018deepinf}, traffic forecasting~\cite{yuspatio}, recommender system~\cite{ying2018graph,van2017graph}, action recognition~\cite{yan2018spatial} and event detection~\cite{nguyen2018graph}. 

Although GNN achieves great performance in several tasks, no work has considered using GNN to model the high-order social influence in social recommender system.

\subsection{Attention Mechanism}
Multiple works~\cite{chen2017attentive,li2017neural,cao2018entive,he2018nais,cheng20183ncf} incorporate the attention mechanism with the implementation of neural networks proposed to improve the performance of recommender systems. Attentive Collaborative Filtering(ACF)~\cite{chen2017attentive} aggregates the item- and component-level implicit feedback with an attention network to get the representation for a multimedia item. Neural Attentive Recommendation Machine(NARM)~\cite{li2017neural} adopts attention mechanism to model the user sequential behavior and capture user's main purpose in a session-based recommendation scenario. Moreover, the AGREE model~\cite{cao2018entive} learns to assign an attention weight for members to solve group recommendation problem. Some works~\cite{xu2018representation, velivckovic2017graph} also employ attention mechanism to improve the performance of graph neural network. GAT~\cite{velivckovic2017graph} utilizes attention mechanism to learn dynamic weights of neighbors. Jump Knowledge Network~\cite{xu2018representation} is proposed to flexibly leverage the neighbor range for each node.

In our work, the attention mechanism is integrated to learn a personalized weight of different order neighbors for each user.

\section{CONCLUSION and FUTURE WORK}
In this work, we propose a novel social recommendation framework HOSR, which integrates the information of high-order neighbors to solve the data sparsity problem. The core of our model is to generate user embedding by performing embedding propagation along high-order social neighbors. Leveraging the graph convolutional layer, we can explicitly model the effect of high-order neighbors into representation framework. Attention mechanism is further employed to leverage the output of different layers, and two dropout strategies are adopted to alleviate overfitting. Experiments performed on two real-world datasets demonstrate the effectiveness of our model and the promise of influence propagation.

In future, we plan to extend our work in three directions: 1) our work focus on learning the representation from social perspective. Multiple works~\cite{yu2018adaptive, guo2015trustsvd} demonstrate jointly considering user's social and interaction relations could improve model performance. Thus, we will consider jointly propagate user and item embedding in future work. 2) In social network, a user may have close and normal friends. we will attempt to utilize attention mechanism to specify attention weights for user-user connections. 3) Experimental results in section \ref{sec:attention} show that using attention to aggregate the output of different layers has little impact on users with sparse social relations. Therefore, we will try to explore a more effective aggregating mechanism to fully utilize the information of different order neighbors.


%

\ifCLASSOPTIONcompsoc
  \section*{Acknowledgments}
\else
  \section*{Acknowledgment}
\fi

The paper was supported by the National Key Research and Development Program (2017YFB0202201), the National Natural Science Foundation of China (61702568, U1711267), the Program for Guangdong Introducing Innovative and Entrepreneurial Teams (2017ZT07X355) and the Fundamental Research Funds for the Central Universities under Grant (17lgpy117).

\ifCLASSOPTIONcaptionsoff
  \newpage
\fi

\bibliographystyle{IEEEtran}
\bibliography{main}


%

%


\begin{IEEEbiography}[{\includegraphics[width=1in,height=1.25in,clip,keepaspectratio]{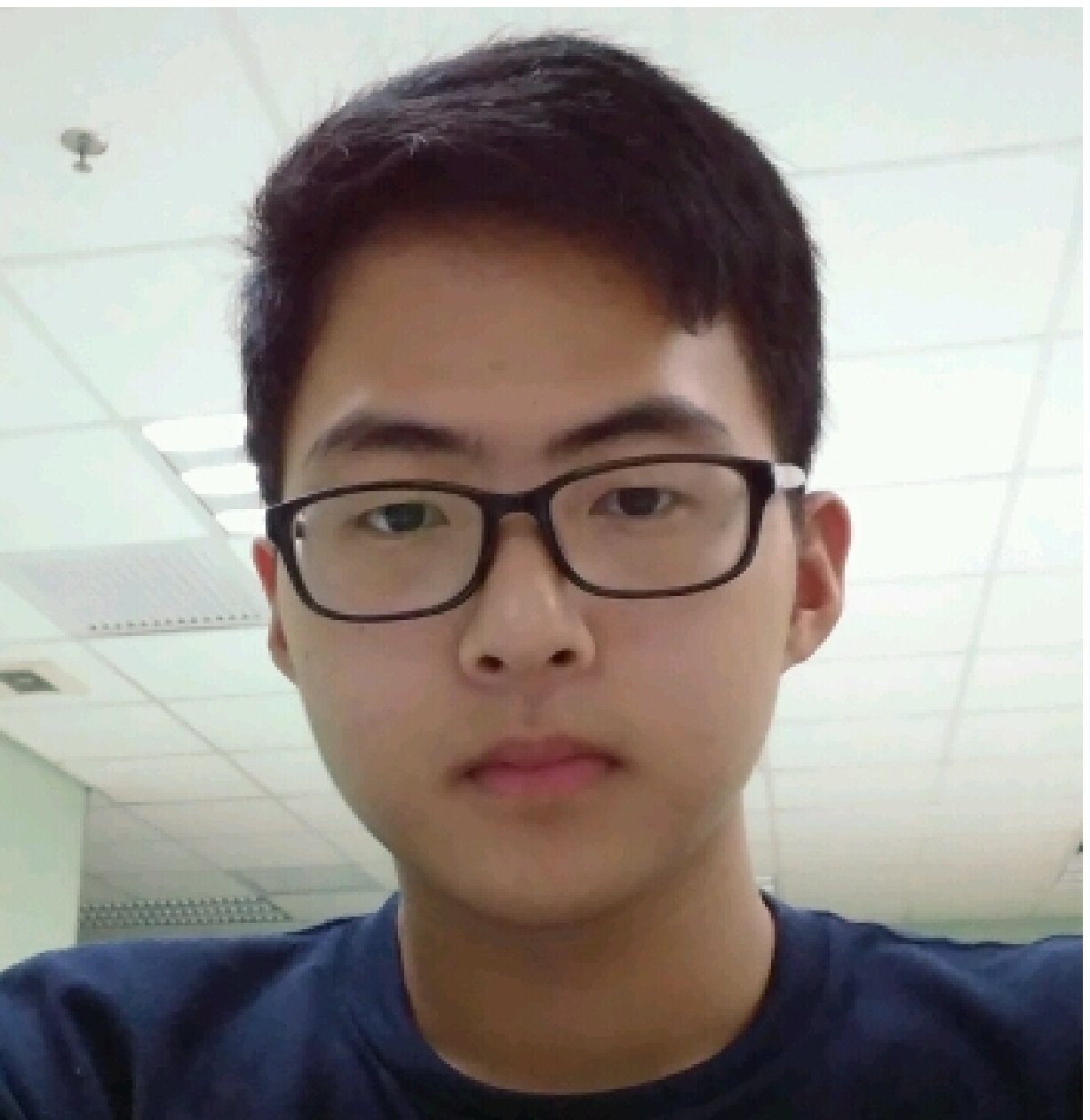}}]{Yang Liu}
received the bachelor's degree at Sun Yat-sen University, Guangzhou, China, in 2019. He is currently pursuing the master's degree with the School of Data and Computer Science, Sun Yat-sen University, Guangzhou, China. His main research interests include recommendation systems, machine learning and data mining techniques.
\end{IEEEbiography}

\begin{IEEEbiography}[{\includegraphics[width=1in,height=1.25in,clip,keepaspectratio]{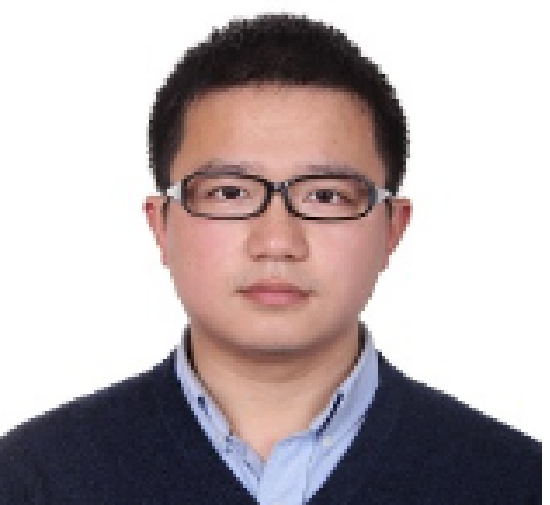}}]{Liang Chen}
received the Ph.D. and bachelor’s degrees from the Advanced Computing and System Laboratory, College of Computer Science and Technology, Zhejiang University, China, in 2015 and 2009, respectively. He is currently a Distinguished Research Fellow with the School of Data and Computer Science, Sun Yat-Sen University, Guangzhou, China. His research areas include services computing, social networks, recommendation systems, and heterogeneous information networks, with a focus on solving traditional challenges via heterogeneous data sources and data mining techniques.
\end{IEEEbiography}

\begin{IEEEbiography}[{\includegraphics[width=1in,height=1.25in,clip,keepaspectratio]{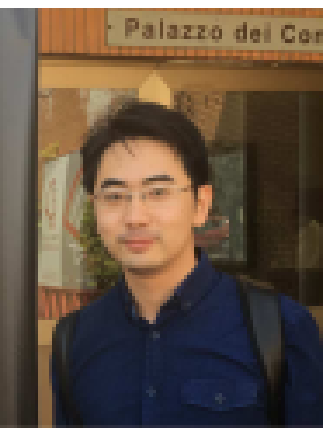}}]{Xiangnan He}
is currently a research fellow with School of Computing, National University of Singapore (NUS). He received his Ph.D. in Computer Science from NUS. His research interests span recommender system, information retrieval, natural language processing and multimedia. His work on recommender system has received the Best Paper Award Honorable Mention in WWW 2018 and SIGIR 2016. Moreover, he has served as the PC member for top-tier conferences including SIGIR, WWW, MM, KDD, WSDM, CIKM, AAAI, and ACL, and the invited reviewer for prestigious journals including TKDE, TOIS, TKDD, TMM, and WWWJ.
\end{IEEEbiography}

\begin{IEEEbiography}[{\includegraphics[width=1in,height=1.25in,clip,keepaspectratio]{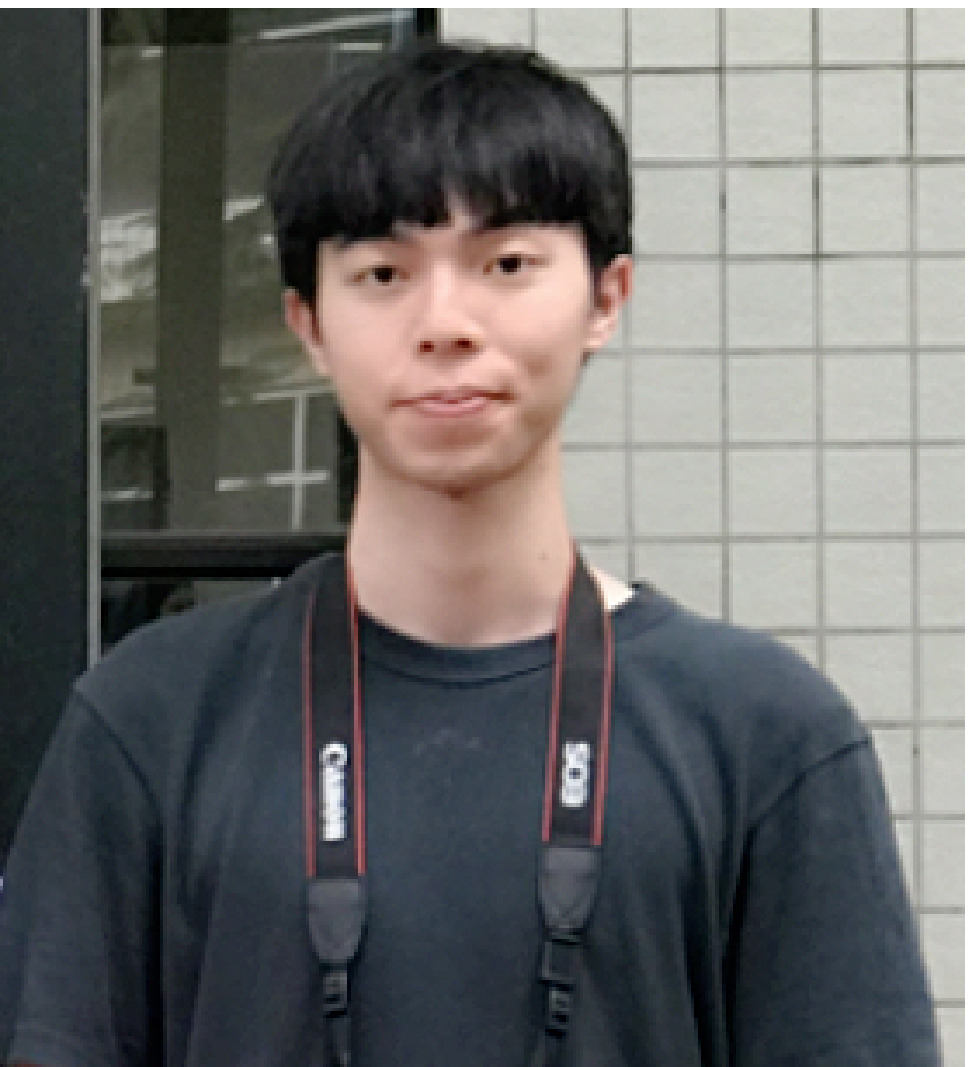}}]{Jiaying Peng}
received the bachelor's degree at South China Normal University, Guangzhou, China, in 2019. He is currently pursuing the master's degree with the School of Data and Computer Science, Sun Yat-sen University, Guangzhou, China. His main research interests include social networks, recommendation systems, machine learning and data mining techniques.
\end{IEEEbiography}

\begin{IEEEbiography}[{\includegraphics[width=1in,height=1.25in,clip,keepaspectratio]{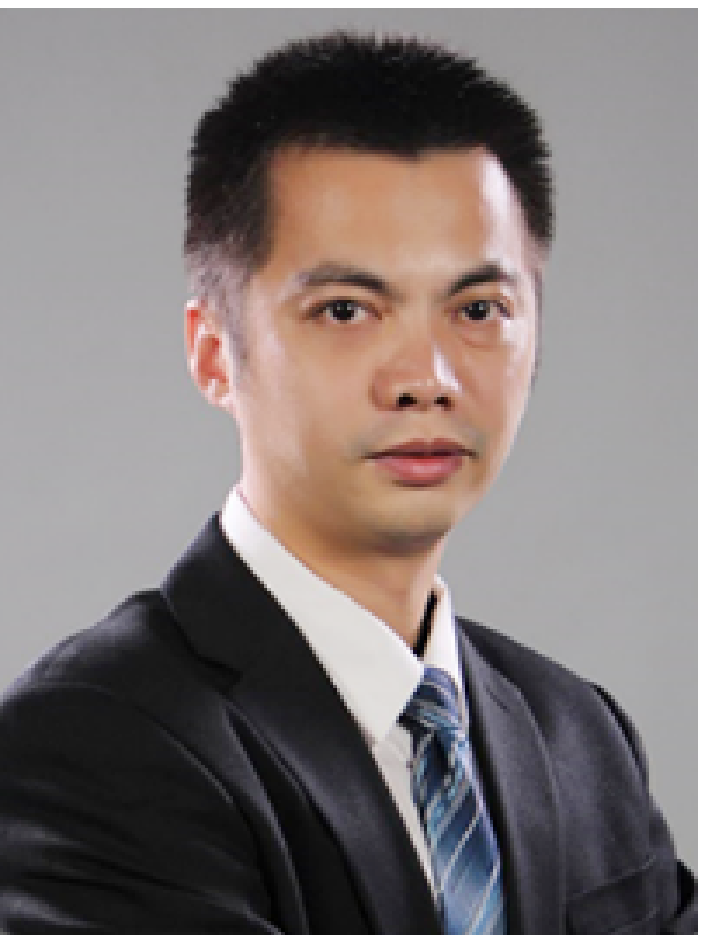}}]{Zibin Zheng}
received the Ph.D. degree from The Chinese University of Hong Kong, in 2011. He is currently a Professor with the School of Data and Computer Science, Sun Yat-sen University, Guangzhou, China. His research interests include services computing, software engineering, and blockchain. He received the ACM SIGSOFT Distinguished Paper Award at the ICSE’10, the Best Student Paper Award at the ICWS’10, and the IBM Ph.D. Fellowship Award.
\end{IEEEbiography}

\begin{IEEEbiography}[{\includegraphics[width=1in,height=1.25in,clip,keepaspectratio]{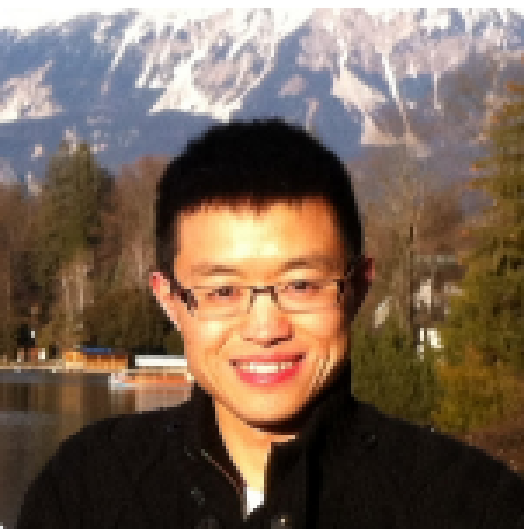}}]{Jie Tang}
is an associate professor with the Department of Computer Science and Technology, Tsinghua University. His main research interests include data mining algorithms and social network theories. He has been a visiting scholar with Cornell University, Chinese University of Hong Kong, Hong Kong University of Science and Technology, and Leuven University. He has published more then 100 research papers in major international journals and conferences including: KDD, IJCAI, AAAI, ICML, WWW, SIGIR, SIGMOD, ACL, Machine Learning Journal, TKDD, and TKDE.
\end{IEEEbiography}




\end{document}